\documentclass[12pt,prd]{revtex4}
\usepackage{amsmath}
\usepackage{graphicx,color}

\newcommand{\arxiv}[1]{{arXiv:#1}}

\begin{document}

\hfill\parbox{4cm}
{
 BNL-94608-2011-JA\\
 YITP-11-6\\
}

\title{
Aoki Phases in the Lattice Gross-Neveu Model \\
with Flavored Mass terms
}

\author{Michael Creutz}
\email{creutz@bnl.gov}
\affiliation{Physics Department, Brookhaven National Laboratory, 
Upton, NY 11973, USA}

\author{Taro Kimura}
\email{kimura@dice.c.u-tokyo.ac.jp}
\affiliation{Department of Basic Science, University of Tokyo,
         Tokyo 153-8902, Japan}

\author{Tatsuhiro Misumi}
\email{misumi@yukawa.kyoto-u.ac.jp}
\affiliation{Yukawa Institute for Theoretical Physics, Kyoto University,
         Kyoto 606-8502, Japan}

\begin{abstract}
We investigate the parity-broken phase structure for staggered and
naive fermions in the Gross-Neveu model as a toy model of QCD. We
consider a generalized staggered Gross-Neveu model including two types
of four-point interactions.  We use generalized mass terms to split
the doublers for both staggered and naive fermions.  The phase
boundaries derived from the gap equations show that the mass
splitting of tastes results in an Aoki phase both in the staggered and
naive cases.  We also discuss the continuum limit of these models and
explore taking the chirally-symmetric limit by fine-tuning a mass
parameter and two coupling constants.  This supports the idea that in
lattice QCD we can derive one- or two-flavor staggered fermions 
by tuning the mass parameter, which are likely to be less expensive than
Wilson fermions in QCD simulation.
\end{abstract}

\maketitle

\newpage

%%%%%%%%%%   Introduction   %%%%%%%%%%

\section{Introduction}
\label{sec:Intro}

Since the pioneering work in Ref.~\cite{AokiP}, the rich
phase structure in the lattice Wilson fermion has been extensively
studied \cite{AokiU1,AH,CreutzW,SS,TNU}. As is well-known \cite{Wil}, 
Wilson fermions bypass the
no-go theorem \cite{NN} and produce a single fermionic degree of
freedom by breaking the chiral symmetry explicitly.  This leads to an
additive mass renormalization and requires fine-tuning of a mass
parameter for a chiral limit. Furthermore at finite lattice spacing,
there emerges a parity-broken phase (Aoki phase) \cite{AokiP}.  The
full phase diagram reflects the masses possessed by each of the
original doublers.  As seen from this fact, the main reason for the
emergence of the parity-broken phase is that the Wilson term gives a
species(taste)-sensitive mass to produce a mass splitting of species
as well as breaking the chiral symmetry.  The understanding of the
parity-broken phase structure is not only useful for simulations with
Wilson fermions, but also gives practical information for the
application of overlap \cite{GW,Neu} and domain-wall \cite{Kap, FuSh}
fermions, both of which are built on the Wilson fermion kernel. Indeed
it is shown in \cite{TN} that the domain-wall fermion also possesses
a complicated parity broken phase diagram for a finite size of the
extra dimension.

On the other hand, no parity-broken phase structure is observed in
staggered fermions \cite{KS,Suss,Sha} with their exact chiral symmetry.  
However things could be changed if we introduce a taste-sensitive mass term, 
which we refer to as a taste-splitting or flavored mass in this paper. 
Adams recently established theoretical foundation of the index theorem 
with staggered fermions \cite{Adams1} and presented a new version of 
the overlap fermion constructed from the staggered kernel \cite{Adams2, Hoel}. 
He introduced a taste-splitting mass term for the spectral flow to detect
the index correctly. This mass term assigns positive and negative
masses to tastes depending on their flavor-chiralities. After these
works the present authors \cite{CKM} successfully defined the index in
the naive and minimally doubled fermions \cite{Kar, Wilc, MD1,MD2} and
presented new versions of overlap fermions by implementing the
flavored mass terms \cite{CreutzPS}.  It is natural to consider
the phase diagram for these fermions with the mass splitting of the
tastes since it is also useful for the practical application of their
overlap versions as well as themselves.

In this paper we study the parity-broken phase structure for naive and 
staggered fermions with the flavored mass terms. We use the
two-dimensional lattice Gross-Neveu models \cite{GN, EN, AH} as toy
models of QCD.  We develop the generalized staggered Gross-Neveu model
including two types of four-point interactions to study the staggered
phase structure.  We solve the gap equations for the large $N$ limit and
obtain the phase boundaries in the $M$-$g^{2}$ plane.  We show the Aoki phase 
exists both in staggered and naive cases reflecting the mass splitting 
of tastes. In the naive cases there are varieties of the phase diagram 
depending on linear combinations of two types of the flavored masses. 
This elucidation of the phase structure can contribute to the practical
application of these fermions and their overlap versions. We also discuss 
the continuum limits of these Gross-Neveu models. We show that we can take 
the chiral continuum limit with the associated number of massless 
fermions by fine-tuning a mass parameter and two coupling constants. It 
indicates that, in Lattice QCD with the staggered fermions with the 
Adams-type \cite{Adams2} or Hoelbling-type \cite{Hoel} flavored masses, 
we can obtain the two- or one-flavor massless fermions in the chiral and 
continuum limit by tuning the mass parameter. They can be less 
expensive than Wilson fermions in lattice QCD simulations.

In Sec.~\ref{sec:NGN} we study the parity broken phase diagram by
using the naive Gross-Neveu model with the flavored mass.  In
Sec.~\ref{sec:SGN} we propose the generalized staggered Gross-Neveu model
and study the phase diagram.  In Sec.~\ref{sec:CL} we investigate the
continuum limit of these models and discuss the first order phase
phase boundaries in the phase diagram. Section \ref{sec:SD} is devoted
to a summary and discussion.

%%%%%%%%%%%%%% Naive Gross-Neveu model %%%%%%%%%%

\section{Naive Gross-Neveu model}
\label{sec:NGN}

In this section we investigate the phase diagram for naive lattice
fermions with flavored mass terms by using the $d=2$ Gross-Neveu
model, which has lots of common features with QCD. Let us begin with
the lattice Gross-Neveu model with the flavored mass term, 
which is given by
\begin{align}
S\,=\,
{1\over{2}}\sum_{n,\mu}\bar{\psi}_{n}\gamma_{\mu}(\psi_{n+\mu}-\psi_{n-\mu})
-{g^{2}\over{2N}}\sum_{n}&[(\bar{\psi}_{n}\psi_{n})^{2}
  +(\bar{\psi}_{n}i\gamma_{5}\psi_{n})^{2}]
\nonumber\\ 
&+\sum_{n,m}\bar{\psi}_{n}(M\delta_{nm}
+{(M_{f})}_{n,m})\psi_{m},
\label{SGNS}
\end{align}
where $\mu$ stands for $\mu=1,2$, $n=(n_1, n_2)$ are the two dimensional 
coordinates and $\psi_{n}$ stands for a $N$-component Dirac fermion field
$(\psi_{n})_{j}$($j=1,2,...,N$).  We note the bilinear
$\bar{\psi}\psi$ means $\sum_{j=1}^{N}\bar{\psi}_{j}\psi_{j}$.
$g^{2}$ corresponds to the 't Hooft coupling.  $M$ is a usual mass
assigning the same mass to species while $(M_{f})_{n,m}$ is a flavored 
mass assigning different masses to them.  
Here we define the two dimensional gamma matrices as
$\gamma_{1}=\sigma_{1}$, $\gamma_{2}=\sigma_{2}$ and
$\gamma_{5}=\sigma_{3}$.  We make all the quantities dimensionless in
this equation.  By introducing auxiliary bosonic fields $\sigma_{n}$,
$\pi_{n}$ we remove the four-point interactions as
\begin{align}
S\,=\, {1\over{2}}\sum_{n,\mu}\bar{\psi}_{n}\gamma_{\mu}(\psi_{n+\mu}&-\psi_{n-\mu})
+\sum_{n,m}\bar{\psi}_{n}(M_{f})_{n,m}\psi_{m}
\nonumber\\
&+{N\over{2g^{2}}}\sum_{n}((\sigma_{n}-M)^{2}+\pi_{n}^{2})
+\sum_{n}\bar{\psi}_{n}(\sigma_{n}+i\gamma_{5}\pi_{n})\psi_{n}.
\label{SGNSsp}
\end{align}
By solving the equations of motion, we show the following relation between these
auxiliary fields and the bilinears of the fermion fields
\begin{align}
\sigma_{n}&=M-{g^{2}\over{N}}\bar{\psi}\psi,
\label{sigma}
\\
\pi_{n}&=-{g^{2}\over{N}}\bar{\psi}i\gamma_{5}\psi.
\label{pi}
\end{align}
These relations indicate how $\sigma$ and $\pi$ stand for the scalar
and pseudo-scalar mesons.  After integrating the fermion fields, the
partition function and the effective action with these auxiliary
fields are given by
\begin{align}
Z\,&=\, \int\prod_{n}(d\sigma_{n}d\pi_{n})e^{-N\,S_{\rm eff}(\sigma,\pi)},
\label{Par}
\\
S_{\rm eff}(\sigma_{n},\pi_{n}) \,&=\, {1\over{2g^{2}}}\sum_{n}((\sigma_{n}-M)^{2}+\pi_{n}^{2})-{\rm Tr}\,\log D_{n,m},
\label{NSeff}
\end{align}
with
\begin{equation}
D_{n,m}=(\sigma_{n}+i\gamma_{5}\pi_{n})\delta_{n.m}+
{\gamma_{\mu}\over{2}}(\delta_{n+\mu,m}-\delta_{n-\mu,m})+(M_{f})_{n,m}.
\label{ND}
\end{equation}
Here ${\rm Tr}$ stands for the trace both for the position and spinor spaces.
As is well-known, the partition function in the Gross-Neveu model is given 
by the saddle point of this effective action in the large $N$ limit.
We denote as $\tilde{\sigma}_{n}$, $\tilde{\pi}_{n}$ solutions satisfying 
the saddle-point conditions
\begin{equation}
{\delta S_{\rm eff}(\sigma_{n},\pi_{n})\over{\delta \sigma_{n}}}\,=\, 
{\delta S_{\rm eff}(\sigma_{n},\pi_{n})\over{\delta \pi_{n}}}\,=\, 0.
\label{NSadEq1}
\end{equation}
Then the partition function is given by
\begin{equation}
Z\,=\, e^{-S_{\rm eff}(\tilde{\sigma},\tilde{\pi})}.
\label{NeffPar}
\end{equation}
By assuming the translational invariance we define the position-independent solutions 
as $\sigma_{0}\equiv\tilde{\sigma}_{0}$ and $\pi_{0}\equiv\tilde{\pi}_{0}$ 
Then we can factorize a volume factor $V=\sum_{n}1$ in the effective action as 
\begin{align}
S_{\rm eff}\,&=\, V\tilde{S}_{\rm eff}(\sigma_{0},\pi_{0}),
\\
\tilde{S}_{\rm eff}(\sigma_{0},\pi_{0})
\,&=\, {1\over{2g^{2}}}((\sigma_{0}-M)^{2}+\pi_{0}^{2})-{1\over{V}}{\rm Tr}\,\log D.
\label{NtilSeff}
\end{align}
We can write ${\rm Tr}\log D$ in a simple form by the Fourier transformation to momentum space
\begin{align}
{\rm Tr}\,\log D \,&=\,V\int{d^{2}k\over{(2\pi)^{2}}}\log[{\rm det}(\sigma_{0}+i\gamma_{5}\pi_{0}+
M_{f}(k)+i\sum_{\mu}\gamma_{\mu}\sin k_{\mu})]
\nonumber\\
&=\, V\int{d^{2}k\over{(2\pi)^{2}}}\log[(\sigma_{0}+M_{f}(k))^{2}+\pi_{0}^{2}+s^{2}],
\label{NI}
\end{align}
with ${\rm det}$ being the determinant in the spinor space 
and $s^{2}=\sum_{\mu}\sin^{2}k_{\mu}$.
$M_{f}(k)$ is the flavored mass represented in momentum space.
Now the saddle-point equations are written as
\begin{align}
{\delta \tilde{S}_{\rm eff}\over{\delta \sigma_{0}}}\,&=\,
{(\sigma_{0}-M)\over{g^{2}}}-2\int{d^{2}k\over{(2\pi)^{2}}}
{\sigma_{0}+M_{f}(k)\over{(\sigma_{0}+M_{f}(k))^{2}+\pi_{0}^{2}+s^{2}}}=0,
\label{Ncond1}
\\ 
{\delta \tilde{S}_{\rm eff}\over{\delta \pi_{0}}}\,&=\,
{\pi_{0}\over{g^{2}}}-2\int{d^{2}k\over{(2\pi)^{2}}}
{\pi_{0}\over{(\sigma_{0}+M_{f}(k))^{2}+\pi_{0}^{2}+s^{2}}}=0.
\label{Ncond2}
\end{align}
In this section we consider two types of the flavored mass for the naive fermion
\begin{align}
M_{f}^{(1)}(k)\,&=\, \cos k_{1} \cos k_{2}, 
\label{M1}
\\
M_{f}^{(2)}(k)\,&=\, {1\over{2}}(\cos k_{1} +\cos k_{2})(1+\cos k_{1} \cos k_{2}). 
\label{M2}
\end{align}
Such mass terms were first introduced in the minimally doubled
fermion by using the point-splitting method \cite{CreutzPS}. Then these
were introduced also for the naive fermion to consider the index theorem
and a new type of overlap fermions \cite{CKM}.  Studying the phase
diagram with these flavored mass terms not only contributes 
to understanding 
the overlap versions but also helps to understand the staggered case
in the next section.  
Here $\sigma_{0}$ and $\pi_{0}$ are determined as 
$\sigma_{0}(M,g^{2})$, $\pi_{0}(M,g^{2})$ from the saddle-point equations 
once the values of $M$ and $g^{2}$ are fixed. 

Let us look into the phase structure with respect to parity symmetry.
The order parameter of this symmetry is $\pi_{0}$, 
which can take zero or non-zero values depending on values of $M$ and $g^{2}$.
Parity symmetry is spontaneously broken for the non-zero cases $\pi_{0}\not=0$.
The phase boundary is determined by imposing $\pi_{0}=0$ on Eq.~(\ref{Ncond1})(\ref{Ncond2})
after the overall $\pi_{0}$ being removed in Eq.~(\ref{Ncond2}) .
Then the conditions for the phase boundary, so-called gap equations, are given by
\begin{align}
{M_{c}\over{g^{2}}}\,&=\,-2\int{d^{2}k\over{(2\pi)^{2}}}
{M_{f}(k)\over{(\sigma_{0}+M_{f}(k))^{2}+s^{2}}},
\label{Ncricon1}
\\
{1\over{g^{2}}}\,&=\,2\int{d^{2}k\over{(2\pi)^{2}}}
{1\over{(\sigma_{0}+M_{f}(k))^{2}+s^{2}}},
\label{Ncricon2}
\end{align}
with $M_{c}$ being the critical value of $M$.
As we will check later, this phase boundary is a second-order critical line. 
Here we derive the parity phase boundary $M_{c}(g^{2})$ as a function of the coupling $g^{2}$
by getting rid of the chiral condensate $\sigma_{0}$ from these equations.
We will calculate the parity phase boundaries for three cases of the flavored masses $M_{f}^{(1)}$, 
$M_{f}^{(2)}$ and $M_{f}^{(1)}+M_{f}^{(2)}$.

\subsection{$M_{f}^{(1)}$}
\label{sec:1} 

\begin{figure}
\centering
\includegraphics[height=5cm]{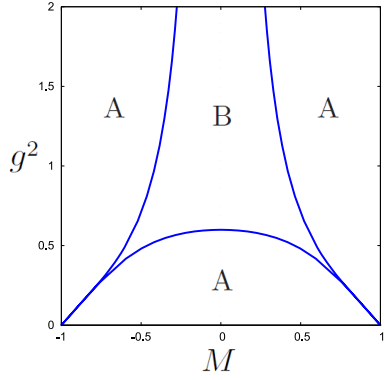} 
\caption{Aoki phase structure for the naive fermion with the flavored mass $M_{f}^{(1)}$. 
The left and right cusps are related to two species ($0,0$)($\pi,\pi$) with $m=1$ 
and the other two ($0,\pi$)($\pi,0$) with $m=-1$ respectively.
A and B stands for parity-symmetric and -broken phases.}
\label{fig1}
\end{figure}

The lattice fermion action with this flavored mass assigns 
the positive mass $m=1$ to two species with the momentum $(0,0)(\pi,\pi)$
and the negative mass $m=-1$ to the other two species with $(0,\pi)(\pi,0)$.
Before calculating $M_{c}(g^{2})$ numerically, we can anticipate the phase structure
from the symmetry of the gap equations. To see this we replace $k_{1}$
by $\pi-k_{1}$ in (\ref{Ncond1}) and (\ref{Ncond2}) for $M_{f}^{(1)}$. 
Then the equations are converted into
\begin{align}
{-\sigma_{0}+M\over{g^{2}}}&=2\int{d^{2}k\over{(2\pi)^{2}}}
{-\sigma_{0}+M_{f}^{(1)}(k)\over{(-\sigma_{0}+M_{f}^{(1)}(k))^{2}+\pi_{0}^{2}+s^{2}}},
\label{NIncond1}
\\ 
{\pi_{0}\over{g^{2}}}&=2\int{d^{2}k\over{(2\pi)^{2}}}
{\pi_{0}\over{(-\sigma_{0}+M_{f}^{(1)}(k))^{2}+\pi_{0}^{2}+s^{2}}}.
\label{NIncond2}
\end{align}
Thus, if ($\sigma_{0}$, $\pi_{0}$) are solutions for ($M$, $g^{2}$),
($-\sigma_{0}$, $\pi_{0}$) are solutions for ($-M$, $g^{2}$).
It also means, if ($M_{c}$, $g^{2}$) is a critical point, ($-M_{c}$, $g^{2}$) too.    
We can anticipate the phase diagram for this case is symmetric about $M=0$.
Now we derive the parity phase boundary $M_{c}(g^{2})$ numerically for $M_{f}^{(1)}(k)=\cos k_{1} \cos k_{2}$.
The phase diagram for this case is depicted in Fig.~{\ref{fig1}}.
A stands for the parity symmetric phase $\pi_{0}=0$ and B for Aoki phase $\pi_{0}\not=0$.
In the large coupling region there are two phase boundaries
while there are four phase boundaries in the weak coupling region.
The left and right cusps correspond to two species ($0,0$)($\pi,\pi$) with the positive mass ($m=1$) 
and the other two ($0,\pi$)($\pi,0$) with the negative mass ($m=-1$) respectively.
It reflects the mass splitting of species given by the flavored mass $M_{f}^{(1)}$.
Here we note we obtain the same result for $-M_{f}^{(1)}$ except that the species ($0,0$)($\pi,\pi$)
live at the right cusp and the other two live at the left. It means the sign of the this flavored mass
is irrelevant for the spectrum of the Dirac operator or the associated Aoki phase. 

\subsection{$M_{f}^{(2)}$}
\label{sec:2} 

The lattice fermion action with this flavored mass assigns 
the positive mass ($m=2$) to one of four species with the momentum $(0,0)$,
zero mass to $(0,\pi)(\pi,0)$ and the negative mass ($m=-2$) 
to $(\pi,\pi)$. To look at the symmetry of the gap equations we replace $k_{\mu}$ 
by $\pi-k_{\mu}$ in (\ref{Ncond1}) and (\ref{Ncond2}) for $M_{f}^{(2)}$. 
Then the equations are converted into
\begin{align}
{-\sigma_{0}+M\over{g^{2}}}&=2\int{d^{2}k\over{(2\pi)^{2}}}
{-\sigma_{0}+M_{f}^{(2)}(k)\over{(-\sigma_{0}+M_{f}^{(2)}(k))^{2}+\pi_{0}^{2}+s^{2}}},
\label{NNIncond1}
\\ 
{\pi_{0}\over{g^{2}}}&=2\int{d^{2}k\over{(2\pi)^{2}}}
{\pi_{0}\over{(-\sigma_{0}+M_{f}^{(2)}(k))^{2}+\pi_{0}^{2}+s^{2}}}.
\label{NNIncond2}
\end{align}
Thus, if ($\sigma_{0}$, $\pi_{0}$) are solutions for ($M$, $g^{2}$),
($-\sigma_{0}$, $\pi_{0}$) are solutions for ($-M$, $g^{2}$).
It also means, if ($M_{c}$, $g^{2}$) is a critical point, ($-M_{c}$, $g^{2}$) too.    
We can anticipate the phase diagram for this case is again symmetric about $M=0$.
Now we derive the parity phase boundary $M_{c}(g^{2})$ numerically for 
$M_{f}^{(2)}(k)=(\cos k_{1}+\cos k_{2})(1+\cos k_{1} \cos k_{2})/2$.
In the large coupling region there are two phase boundaries
while there are six phase boundaries in the weak coupling region.
The three cusps correspond to one of four species ($0,0$) with $m=2$,
two of them $(0,\pi)(\pi,0)$ with $m=0$ and the other one $(\pi,\pi)$ 
with $m=-2$ respectively from the left. 
It reflects the mass splitting of species given by the flavored mass $M_{f}^{(2)}$.

\begin{figure}
\centering
\includegraphics[height=5cm]{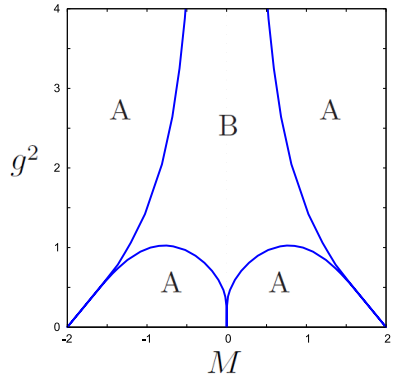} 
\caption{Aoki phase structure for the naive fermion with the flavored mass $M_{f}^{(2)}$. 
The three cusps correspond to ($0,0$) with $m=2$,
$(0,\pi)(\pi,0)$ with $m=0$ and $(\pi,\pi)$ 
with $m=-2$ respectively from the left.}
\label{fig2}
\end{figure}

\subsection{$M_{f}^{(1)}+M_{f}^{(2)}$}
\label{sec:3} 

\begin{figure}
\centering
\includegraphics[height=5cm]{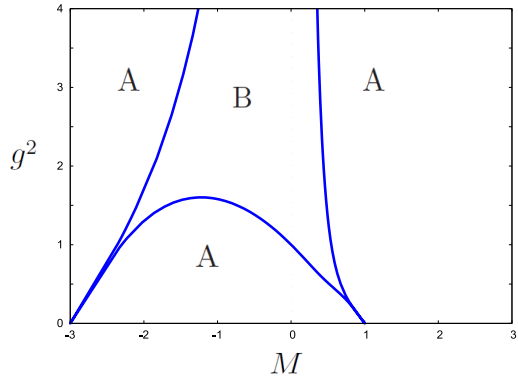} 
\caption{Aoki phase structure for the naive fermion with the flavored mass $M_{f}^{(1)}+M_{f}^{(2)}$. 
The left and right cusps correspond to ($0,0$) with $m=3$ 
and ($0,\pi$)($\pi,0$)($\pi,\pi$) with $m=-1$ respectively.}
\label{fig3}
\end{figure}

The fermion action with this flavored mass assigns 
the positive mass $m=3$ to one of species with the momentum $(0,0)$
and the negative mass $m=-1$ to the other three species with $(0,\pi)(\pi,0)(\pi,\pi)$.
Here we cannot find any relevant symmetry in the gap equations.   
Thus we can anticipate the phase diagram for this case is not symmetric.
Now we calculate $M_{c}(g^{2})$ numerically for 
$M_{f}^{(1)}+M_{f}^{(2)}(k)=\cos k_{1} \cos k_{2}+(\cos k_{1}+\cos k_{2})(1+\cos k_{1} \cos k_{2})/2$.
The result of the phase diagram is depicted in Fig.~{\ref{fig3}}.
It is obvious that it is not symmetric about $M=0$.
In the large coupling region there are two phase boundaries
while there are four phase boundaries in the weak coupling region.
The left and right cusps correspond to one of species ($0,0$) with $m=3$ 
and the other three ($0,\pi$)($\pi,0$)($\pi,\pi$) with $m=-1$ respectively.
It reflects the mass splitting of species given by the flavored mass $M_{f}^{(1)}+M_{f}^{(2)}$.
Now we can easily modify the phase diagram 
by choosing the linear combination of $M^{(1)}_{f}$ and $M_{f}^{(2)}$.

We expect these results are qualitatively 
similar to the phase diagram of the $d=4$ fermion actions with the 
Non-abelian gauge field like QCD except for the number of species
associated with each cusp. 
In the end of this section we check the mass of the $\pi$-meson becomes 
zero on the critical line $M_{c}(g^{2})$.
As is well-known, the correlation length gets infinitely large 
in the vicinity of the second and higher phase boundaries,
which leads to massless dynamical degrees of freedom.
In the case of lattice QCD with chiral-symmetry-broken fermions like Wilson fermion, 
the fine-tuning of the mass parameter to the 2nd order phase boundary leads 
to the chiral limit with massless quarks and massless pions regarded as Goldstone bosons
due to the spontaneous chiral symmetry breaking.
Thus it is quite important to verify it.
We can show the mass of $\pi_{n}$ becomes zero on the phase boundaries as 
\begin{align}
m_{\pi}^{2}\, &\propto\, 
\langle {\delta^{2}S_{\rm eff}\over{\delta \pi_{n}} \delta \pi_{m}} \rangle |_{M=M_{c}} 
=V{\delta^{2}\tilde{S}_{\rm eff}\over{\delta^{2}\pi_{0}^{2}}}|_{M=M_{c}}
\nonumber\\
&=V\Big[ {1\over{g^{2}}}
-2\int{d^{2}k\over{(2\pi)^{2}}}{1\over{(\sigma_{0}+M_{f}(k))^{2}+\pi_{0}^{2}+s^{2}}}
\nonumber\\
&\,\,\,\,\,\,\,\,\,\,\,\,-(2\pi_{0}^{2})\int{d^{2}k\over{(2\pi)^{2}}}{1\over{((\sigma_{0}+M_{f}(k))^{2}+\pi_{0}^{2}+s^{2})^{2}}} \Big]|_{\pi_{0}=0}
\nonumber\\
&=0.
\end{align}
The zero mass of the pion means the phase boundary we derived is the second-order
critical line. We can also check the order of the phase boundaries by depicting the
potential for $\sigma_{0}$ and $\pi_{0}$ as we will discuss in Sec.~\ref{sec:CL}.

%%%%%%%%%%% Staggered Gross-Neveu model %%%%%%%%%%

\section{Staggered Gross-Neveu model}
\label{sec:SGN}

\begin{figure}
\centering
\includegraphics[height=5cm]{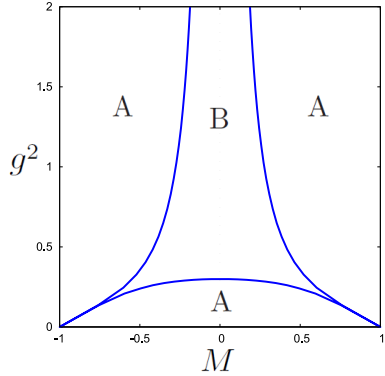} 
\caption{Aoki phase structure for the staggered fermion with 
the Adams-type flavored mass $\Gamma_{5}\Gamma_{55}$. 
The left and right cusps correspond to one of two tastes with $m=1$ 
and the other with $m=-1$. A stands for a parity symmetric phase and 
B for Aoki phase.}
\label{fig4}
\end{figure}

In this section we investigate the phase diagram for staggered fermions with
the Adams-type flavored mass term by using the $d=2$ Gross-Neveu model.
To study the parity broken phase structure we propose the generalized staggered Gross-Neveu model 
with the $\gamma_{5}$-type 4-point interaction, which is given by
\begin{align}
S\,=\, {1\over{2}}&\sum_{n,\mu}\eta_{\mu}\bar{\chi}_{n}(\chi_{n+\mu}-\chi_{n-\mu})
+\sum_{n}\bar{\chi}_{n}(M+M_{f})\chi_{n}
\nonumber\\
&-{g^{2}\over{2N}}\sum_{\mathcal{N}}\Big[(\sum_{A}\bar{\chi}_{2\mathcal{N}+A}\,\chi_{2\mathcal{N}+A})^{2}+(\sum_{A}i(-1)^{A_{1}
+A_{2}}\bar{\chi}_{2\mathcal{N}+A}\,\chi_{2\mathcal{N}+A})^{2}\Big],
\label{Ssta}
\end{align}
where we define two-dimensional coordinates as $n=2\mathcal{N}+A$ with the sublattice $A=(A_{1},A_{2})$
($A_{1,2}=0,1$). $\chi_{n}$ is a one-component fermionic field.
$(-1)^{A_{1}+A_{2}}$ corresponds to the natural definition of $\gamma_{5}$ for this fermion
which is expressed as $\Gamma_{55}=\gamma_{5}\otimes\gamma_{5}$ in the spinor-taste expression. 
$\eta_{\mu}=(-1)^{n_{1}+...+n_{\mu-1}}$ corresponds to $\gamma_{\mu}$.
As the flavored mass term we choose the Adams-type one, which is given by 
\begin{equation}
M_{f}=\Gamma_{5}\Gamma_{55}\sim{\bf 1}\otimes\gamma_{5}+O(a)
\label{Msta}
\end{equation} 
with the following chirality matrix $\Gamma_{5}$
\begin{align}
\Gamma_{5}&=-i\eta_{1}\eta_{2}\sum_{\rm sym}C_{1}C_{2},
\label{G5}
\\
C_{\mu}&={1\over{2}}(T_{\mu}+T_{-\mu})
\label{C}
\end{align}
where $T_{\mu}$ is the usual translation operator.
(The chirality matrix in general dimensions is defined 
as $\Gamma_{5}\equiv-(i)^{d/2} \eta_{1}\cdot\cdot\cdot\eta_{d}\sum_{\rm sym}C_{1}\cdot\cdot\cdot C_{d}$.)
This mass term assigns the positive mass ($m=+1$) to one taste and the negative mass ($m=-1$)
to the other depending on $\pm$ eigenvalues for $\Gamma_{5}\Gamma_{55}$ which we call the flavor-chirality.
With bosonic auxiliary fields $\sigma_{\mathcal{N}}$, $\pi_{\mathcal{N}}$, the action is rewritten as
\begin{align}
S\,=\, {1\over{2}}\sum_{n,\mu}&\eta_{\mu}\bar{\chi}_{n}(\chi_{n+\mu}-\chi_{n-\mu})
+\sum_{n}\bar{\chi}_{n}M_{f}\chi_{n}
\nonumber\\
&+{N\over{2g^{2}}}\sum_{\mathcal{N}}((\sigma_{\mathcal{N}}-M)^{2}+\pi_{\mathcal{N}}^{2})
+\sum_{\mathcal{N},A}\bar{\chi}_{2\mathcal{N}+A}(\sigma_{\mathcal{N}}+i(-1)^{A_{1}+A_{2}}\pi_{\mathcal{N}})\chi_{2\mathcal{N}+A},
\label{StS}
\end{align}
After integrating the fermion field, the partition function and the effective action 
with these auxiliary fields are given by
\begin{align}
Z\,&=\, \int (\mathcal{D}\sigma_{\mathcal{N}}\mathcal{D}\pi_{\mathcal{N}})e^{-N\,S_{\rm eff}(\sigma,\pi)},
\label{StZ}
\\
S_{\rm eff} \,&=\, {1\over{2g^{2}}}\sum_{\mathcal{N}}(\sigma_{\mathcal{N}}^{2}+\pi_{\mathcal{N}}^{2})-{\rm Tr}\,\log D,
\label{StSeff}
\end{align}
with
\begin{equation}
D_{n,m}=(\sigma_{\mathcal{N}}+i(-1)^{A_{1}+A_{2}}\pi_{\mathcal{N}})\delta_{n,m}+
{\eta_{\mu}\over{2}}(\delta_{n+\mu, m}-\delta_{n-\mu, m})+(M_{f})_{n,m}.
\label{StDOp}
\end{equation}
The process from (\ref{NSadEq1}) to (\ref{NtilSeff}) in the case of the naive fermion
is common with this staggered case.
We again denote as $\sigma_{0}$ and $\pi_{0}$ the position-independent solutions of
the saddle-point equations.
In this case, however, it is not straightforward to derive the ${\rm Tr}\log D$ 
with the Dirac operator (\ref{StDOp}) in the effective action Eq.~(\ref{NtilSeff}). 
In order to estimate this trace logarithm we first obtain the determinant of the Dirac operator 
in the sublattice space, which means the determinant in the spinor and taste spaces.
Here we express the sublattice structure as a multiplet field 
$\tilde{\chi}_{\mathcal{N}}$ composed of the four one-component fields as
\begin{equation} 
\tilde{\chi}_{\mathcal{N}}=
\left( 
\begin{array}{ccc}
\chi_{\rm i} \\
\chi_{\rm ii} \\
\chi_{\rm iii} \\
\chi_{\rm iv} \\
\end{array} 
\right)
\label{Stsubmul}
\end{equation}
where we mean ${\rm i}=2\mathcal{N}$, ${\rm ii}=2\mathcal{N}+(1,0)$, ${\rm iii}=2\mathcal{N}+(0,1)$
and ${\rm iv}=2\mathcal{N}+(1,1)$.
Now let us estimate the trace term
\begin{equation}
{\rm Tr}\,\log D = V\int{dk^{2}\over{(2\pi)^{2}}}\log{\rm det}((D(k))_{ab},
\label{Sttrace}
\end{equation}
where $a,b$ stand for the index of the four sublattices
running from ${\rm i}$ to ${\rm iv}$. Here ${\rm det}$ means the determinant with respect
to the sublattice.
The Dirac operator is given by
\begin{align}
(D(k))_{ab}=
\sigma_{0}\delta_{ab}
&+\left( 
\begin{matrix}
+ &   &   &  \\
  & - &   &  \\
  &   & - &  \\
  &   &   & +\\
\end{matrix} 
\right)i\pi_{0}
\nonumber\\
&+i\left( 
\begin{matrix}
 &   &   & + \\
 &  &  + &   \\
  & -  &  &   \\
-  &   &   & \\
\end{matrix} 
\right)\cos{k_{1}\over{2}}\cos{k_{2}\over{2}}
\nonumber\\
&+\left( 
\begin{matrix}
0 & i\sin{k_{1}\over{2}}  & i\sin{k_{2}\over{2}}  & 0 \\
i\sin{k_{1}\over{2}}  & 0 & 0  & -i\sin{k_{2}\over{2}} \\
i\sin{k_{2}\over{2}}  & 0 & 0 &  i\sin{k_{1}\over{2}}\\
0  & -i\sin{k_{2}\over{2}}  & i\sin{k_{1}\over{2}}  & 0\\
\end{matrix} 
\right).
\label{StDab}
\end{align}
Then ${\rm det} D$ is given by
\begin{align}
{\rm det} (D(k))_{ab} &= (\sigma_{0}^{2}+\pi_{0}^{2}+s^{2})^{2}
-2c_{1}^{2}c_{2}^{2}(\sigma_{0}^{2}-\pi_{0}^{2}-s^{2})
+c_{1}^{4}c_{2}^{4}
\nonumber\\
&=((\sigma_{0}+c_{1}c_{2})^{2}+\pi_{0}^2+s^{2})((\sigma_{0}-c_{1}c_{2})^2+\pi_{0}^2+s^{2}),
\label{stdet}
\end{align}
where $s_{\mu}=\sin k_{\mu}/2$, $s^{2}=\sum_{\mu}s_{\mu}^{2}$, 
$c_{\mu}=\cos k_{\mu}/2$.
It is notable that this determinant is expressed by the product of the two determinants of the
naive fermions with the flavored mass $\pm M_{f}^{(1)}(k_{\mu}/2)$.
Now we can explicitly write the saddle-point conditions satisfied by $\sigma_{0}$ and $\pi_{0}$ as
\begin{align} 
{\sigma_{0}-M\over{g^{2}}}&=4\int{dk^{2}\over{(2\pi)^2}}{\sigma_{0}(\sigma_{0}^{2}+\pi_{0}^{2}+s^{2})
-c_{1}^{2}c_{2}^{2}\sigma_{0}\over{((\sigma_{0}+c_{1}c_{2})^{2}+\pi_{0}^2+s^{2})((\sigma_{0}-c_{1}c_{2})^2+\pi_{0}^2+s^{2})}},
\label{StspCon1}
\\
{\pi_{0}\over{g^{2}}}&=4\int{dk^{2}\over{(2\pi)^2}}{\pi_{0}(\sigma_{0}^{2}+\pi_{0}^{2}+s^{2})
+c_{1}^{2}c_{2}^{2}\pi_{0}\over{((\sigma_{0}+c_{1}c_{2})^{2}+\pi_{0}^2+s^{2})((\sigma_{0}-c_{1}c_{2})^2+\pi_{0}^2+s^{2})}}.
\label{StspCon2}
\end{align}
By multiplying $-1$ to the first equation, we see
($-\sigma_{0}$, $\pi_{0}$) are solutions for ($-M$, $g^{2}$)
if ($\sigma_{0}$, $\pi_{0}$) are solutions for ($M$, $g^{2}$).
It also means, if ($M_{c}$, $g^{2}$) is a critical point, ($-M_{c}$, $g^{2}$) too.    
The phase diagram will be symmetric about $M=0$.
The parity phase boundary $M_{c}(g^{2})$ in this case is derived by imposing 
$\pi_{0}=0$ in (\ref{StspCon1})(\ref{StspCon2}) after the overall $\pi_{0}$ being removed in the second one.
Then the gap equations are given by
\begin{align} 
{M_{c}\over{g^{2}}}&=4\int{dk^{2}\over{(2\pi)^2}}{2c_{1}^{2}c_{2}^{2}\sigma_{0}
\over{((\sigma_{0}+c_{1}c_{2})^{2}+\pi_{0}^2+s^{2})((\sigma_{0}-c_{1}c_{2})^2+\pi_{0}^2+s^{2})}},
\\
{1\over{g^{2}}}&=4\int{dk^{2}\over{(2\pi)^2}}{\sigma_{0}^{2}+s^{2}+c_{1}^{2}c_{2}^{2}
\over{((\sigma_{0}+c_{1}c_{2})^{2}+\pi_{0}^2+s^{2})((\sigma_{0}-c_{1}c_{2})^2+\pi_{0}^2+s^{2})}}.
\end{align}
By removing $\sigma_{0}$ in these equations, we derive the phase boundary $M_{c}(g^{2})$.
The result is shown in Fig.~\ref{fig4}.

Here again A stands for the parity symmetric phase ($\pi_{0}=0$) and B for Aoki phase ($\pi_{0}\not=0$).
In the large coupling region there are two phase boundaries
while there are four phase boundaries in the weak coupling region.
The left cusp corresponds to one of two tastes with $m=1$, and the right corresponds to 
the other taste with $m=-1$.
Thus the phase diagram reflects the mass splitting of tastes given 
by the Adams-type flavored mass.
We also check the pion mass becomes zero on the second order phase boundary as
\begin{equation}
m_{\pi}^{2}\, \propto\, 
\langle {\delta^{2}S_{\rm eff}\over{\delta \pi_{n}} \delta \pi_{m}} \rangle |_{M=M_{c}} 
=V{\delta^{2}\tilde{S}_{\rm eff}\over{\delta^{2}\pi_{0}^{2}}}|_{M=M_{c}}=0.
\end{equation}
Now let us consider the parity phase structure in the $d=4$ QCD
with the staggered fermion with this flavored mass.
Considering the case of the Wilson fermion we can speculate it is 
qualitatively similar to our result for the $d=2$ Gross-Neveu model 
except the number of species associated with each cusp. 
In the four dimension, four tastes
in the staggered fermion with the Adams-type flavored mass 
split into two with positive mass and the other two
with negative mass depending on their flavor-chiralities. 
Thus each of the cusps in the phase diagram will correspond to 
two tastes. If we consider another type of the flavored mass term proposed in
\cite{Hoel}, the four tastes are split into one with positive mass,
two with zero mass and the other with negative mass.
If we can take the chiral and continuum limit around the cusps, 
we obtain the two- or one-flavor staggered fermions 
with tuning only the mass parameter,
which will be numerically faster than Wilson fermion.
Thus the question here is whether we can take 
the massless continuum limit. 
We will discuss this point in the next section with starting with
the case of the naive fermion.

%%%%%%%%   Continuum limit   %%%%%%%%%%%%%

\section{Continuum limit}
\label{sec:CL}

In this section we discuss the continuum limit of the naive and 
staggered Gross-Neveu models with the flavored masses discussed 
in Sec.~\ref{sec:NGN} and Sec.~\ref{sec:SGN}.
This analysis gives us important informations on the continuum limit 
of the $d=4$ QCD with these fermions. As is well-known, the chiral 
symmetry is realized in the effective potential of the Gross-Neveu model
as the $O(2)$ rotational symmetry about $\sigma_{0}$ and $\pi_{0}$.
The purpose here is to figure out the fine-tuned values of the mass 
and couplings to recover this symmetry for $a\to0$. We note in order 
to take the chiral and continuum limit in this model, we need 
to introduce two independent couplings $g_{\sigma}^{2}$ and $g_{\pi}^{2}$ 
\cite{AH} as we will see later. The strategy is to expand the fermion 
determinant in the effective potential with respect to the lattice 
spacing $a$ following the process in \cite{AH}.

We first consider the case of the naive fermion with one of the flavored
masses $M_{f}^{(1)}=\cos k_{1} \cos k_{2}$. The effective potential in 
this case with the lattice spacing being explicit is given by
\begin{align}
\tilde{S}_{\rm eff}(\sigma_{0},\pi_{0})
\,&=\, {(\sigma_{0}-M)^2\over{2g_{\sigma}^{2}}}+{\pi_{0}^{2}\over{2}g_{\pi}^2}-I,
\\
I\,&=\, \int^{\pi/a}_{-\pi/a}{d^{2}k\over{(2\pi)^{2}}}\log[(\sigma_{0}+{1\over{a}}\cos k_{1}a\cos k_{2}a)^{2}
+\pi_{0}^{2}+\sum_{\mu}{\sin ^2 k_{\mu}a\over{a^2}}].
\label{oriSeff}
\end{align}
Now we divide the terms in the determinant $I$ into $\mathcal{O}(1/a^2)$ and $\mathcal{O}(1/a)$ parts as
\begin{align} 
I(D_{0},D_{1})&= \int^{\pi/a}_{-\pi/a}{d^{2}k\over{(2\pi)^{2}}}\log[D_{0}+D_{1}],
\label{IDD}
\\
D_{0}&\equiv \sum_{\mu}{\sin ^2 k_{\mu}a\over{a^2}}+(\sigma_{0}-{\alpha\over{a}})^{2}+\pi_{0}^2+\Big({\alpha+\cos k_{1}a\cos k_{2}a\over{a}} \Big)^2 .
\label{D0}
\\
D_{1}&\equiv 2(\sigma_{0}-{\alpha\over{a}})\Big({\alpha+\cos k_{1}a\cos k_{2}a\over{a}} \Big).
\label{D1}
\end{align}
where we introduce a constant $\alpha$ since there is 
arbitrariness about how to divide the terms into $\mathcal{O}(1/a^2)$ and $\mathcal{O}(1/a)$ parts.
This is determined by which cusp you choose in Fig.~\ref{fig1}, 
or equivalently which species you want to make massless in the continuum limit. 
Here we fix $\alpha=-1$ which is related to the left cusp or the continuum limit 
with the massless species $(0,0)$ and $(\pi,\pi)$. 
(With $\alpha=1$ we can discuss the other cusp
while we will discuss $\alpha=0$ in the end of this section.)
Here we use the shifted definition of $\sigma_{0}$ as $\sigma_{0}+1/a\,\to\,\sigma_{0}$
for simplicity for a while. Then the effective potential with this shift is given by
\begin{align}
\tilde{S}_{\rm eff}(\sigma_{0},\pi_{0})
\,&=\, {(\sigma_{0}-(M+1/a))^2\over{2g_{\sigma}^{2}}}+{\pi_{0}^{2}\over{2}g_{\pi}^2}-I(D_{0},D_{1}).
\\
D_{0}&= \sum_{\mu}{\sin ^2 k_{\mu}a\over{a^2}}+\sigma_{0}^{2}+\pi_{0}^2+\Big({-1+\cos k_{1}a\cos k_{2}a\over{a}} \Big)^2 .
\\
D_{1}&= 2\sigma_{0}\Big({-1+\cos k_{1}a\cos k_{2}a\over{a}} \Big).
\end{align}
We expand $I$ by $D_{1}/D_{0}$ or equivalently by the lattice spacing $a$,
\begin{align}
I&=I_{0}+\sum_{n=1}I_{n}, 
 \label{naive_int}
\\
I_{0}&=\int^{\pi/a}_{-\pi/a}{d^{2}k\over{(2\pi)^{2}}}\log D_{0},
 \label{naive_int0}
\\
I_{n}&=-{(-1)^n \over{n}} \int^{\pi/a}_{-\pi/a}{d^{2}k\over{(2\pi)^{2}}} {D_{1}^{n}\over{D_{0}^{n}}}\,\,\,\,\,\,\,(n\geq 1 ),
\nonumber\\
&=-{(-1)^n \over{n}}(2\sigma_{0})^{n}a^{n-2}
\nonumber\\
&\,\,\,\,\, \times \int^{\pi}_{-\pi}{d^{2}\xi\over{(2\pi)^2}}
{(-1+\cos \xi_{1} \cos \xi_{2})^{n}\over{(\sum_{\mu}\sin^{2}\xi_{\mu}+
(-1+\cos \xi_{1} \cos \xi_{2})^2+a^{2}(\sigma_{0}^{2}+\pi_{0}^{2}))^n}},
 \label{naive_int1}
\end{align} 
where we introduce the dimensionless momentum $\xi_{\mu}=k_{\mu}a$.
For $a\to 0$, only the $I_{0}$, $I_{1}$ and $I_{2}$ remains nonzero.
$I_{0}(a\to 0)$, $I_{1}(a\to 0)$ and $I_{2}(a\to 0)$ are given by 
\begin{align} 
I_{0}(a\to 0)&=\tilde{C}_{0}(\sigma_{0}^{2}+\pi_{0}^2)-{1\over{2\pi}}(\sigma_{0}^{2}+\pi_{0}^2)\log{a^{2}(\sigma_{0}^{2}+\pi_{0}^2)\over{e}}
\,\,\,\,\,\,\,\,\,\,\,(\tilde{C}_{0}=0.367),
 \label{naive_int2} \\
I_{1}(a\to 0)&={2\sigma_{0}\over{a}}C_{1}\,\,\,\,\,\,\,\,\,\,\,(C_{1}=-0.446),
 \label{naive_int3} \\
I_{2}(a\to
 0)&=-2\sigma_{0}^{2}C_{2}\,\,\,\,\,\,\,\,\,\,\,\,(C_{2}=0.201).
 \label{naive_int4}
\end{align}
From here we basically do not care about the $\mathcal{O}(a)$ corrections.
Here we show the explicit values of $\tilde{C}_{0}$, $C_{1}$ and $C_{2}$ 
since they will be essential for the discussion later.
The details of the calculations are shown in Appendix \ref{sec:NeffC}.
Now let us discuss the continuum limit of this theory.
Including all the nonzero contributions for $a\to 0$, the effective potential is given by
\begin{align}
\tilde{S}_{\rm eff}&=-\Big({M+1/a\over{g_{\sigma}^2}}+{2\over{a}}C_{1}\Big)\sigma_{0}
+\Big( {1\over{2g_{\pi}^2}}-\tilde{C}_{0} + {1\over{2\pi}}\log a^{2} \Big)\pi_{0}^{2}
\nonumber\\
&+\Big( {1\over{2g_{\sigma}^2}}-\tilde{C}_{0}+ 2C_{2} + {1\over{2\pi}}\log a^{2} \Big)\sigma_{0}^{2}
+{1\over{2\pi}}(\sigma_{0}^{2}+\pi_{0}^2)\log{\sigma_{0}^{2}+\pi_{0}^2\over{e}}.
\label{renS}
\end{align}
This indicates we need two independent couplings $g_{\sigma}^{2}$, $g_{\pi}^{2}$ to
recover the $O(2)$ symmetry toward the continuum limit. 
In addition, getting rid of the $\sigma_{0}$ linear
term leads to the massless limit. 
Then the natural fine-tuned parameters for the chirally symmetric continuum 
limit without $\mathcal{O}(a)$ corrections are given by
\begin{align}
M&=-{2g_{\sigma}^{2}\over{a}}C_{1}-1,
\label{MC1}
\\
g_{\pi}^{2}&={g_{\sigma}^{2}\over{4C_{2}g_{\sigma}^{2}+1}},
\label{gC2}
\end{align}
where Eq.~(\ref{MC1}) is obtained by imposing the coefficient 
of $\sigma_{0}$ and Eq.~(\ref{gC2}) is given by
imposing the coefficients of $\sigma_{0}^{2}$ and $\pi_{0}^{2}$ coincide.
To consider a renormalized theory with the chiral symmetry
we introduce the scale parameter ($\Lambda$-parameter) as
\begin{align}
\Lambda a = {\rm exp}\left[\pi \tilde{C}_{0}-2\pi C_{2}-{\pi\over{2 g_{\sigma}^{2}}}\right].
\label{lambda}
\end{align}
With the natural fine-tuning (\ref{gC2}), this definition of $\Lambda$ leads to
the coupling renormalization including $a$ given by
\begin{align}
{1\over{2g_{\sigma}^{2}}}&=\tilde{C}_{0}- 2C_{2} + {1\over{2\pi}}\log \left( {1\over{\Lambda^{2}a^{2}}}\right),
\label{arengs}
\\
{1\over{2g_{\pi}^{2}}}&=\tilde{C}_{0} + {1\over{2\pi}}\log \left( {1\over{\Lambda^{2}a^{2}}}\right).
\label{arengp}
\end{align}
Here we need to keep $\Lambda$ finite when we take the continuum limit $a\to 0$.
Then the renormalized effective potential with the chiral symmetry in the continuum limit is given by
\begin{equation}
\tilde{S}_{\rm eff}={1\over{2\pi}}(\sigma_{0}^{2}+\pi_{0}^{2})\log{\sigma_{0}^{2}+\pi_{0}^{2}\over{e\Lambda^{2}}}
\label{renaS}
\end{equation}
We note the fine-tuned point $(M(g_{\sigma}^{2}), g_{\pi}^{2}(g_{\sigma}^{2}))$ 
in (\ref{MC1})(\ref{gC2}) specifies the line along which the continuum limit should be taken.
At the minimum of this potential $\sigma_{0}$ has a nonzero value, which
corresponds to the spontaneous chiral symmetry breaking.

\begin{figure}
\centering
\includegraphics[height=5cm]{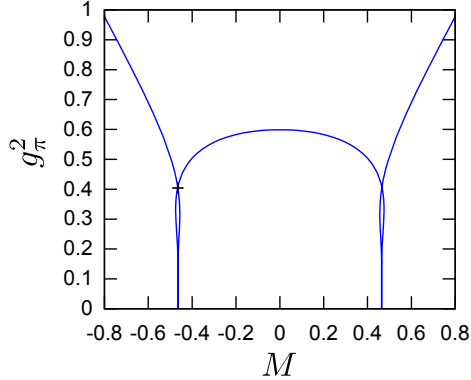} 
\caption{The naively derived phase boundary $M(g_{\pi}^{2})$ 
for the naive fermion with $M_{f}^{(1)}$ 
with $g_{\sigma}^{2}=0.6$. The fine-tuned point $(-0.464, 0.404)$
as a crosspoint is located near the self-crossing point.}
\label{NMgg}
\end{figure}

\begin{figure}
\centering
\includegraphics[height=5cm]{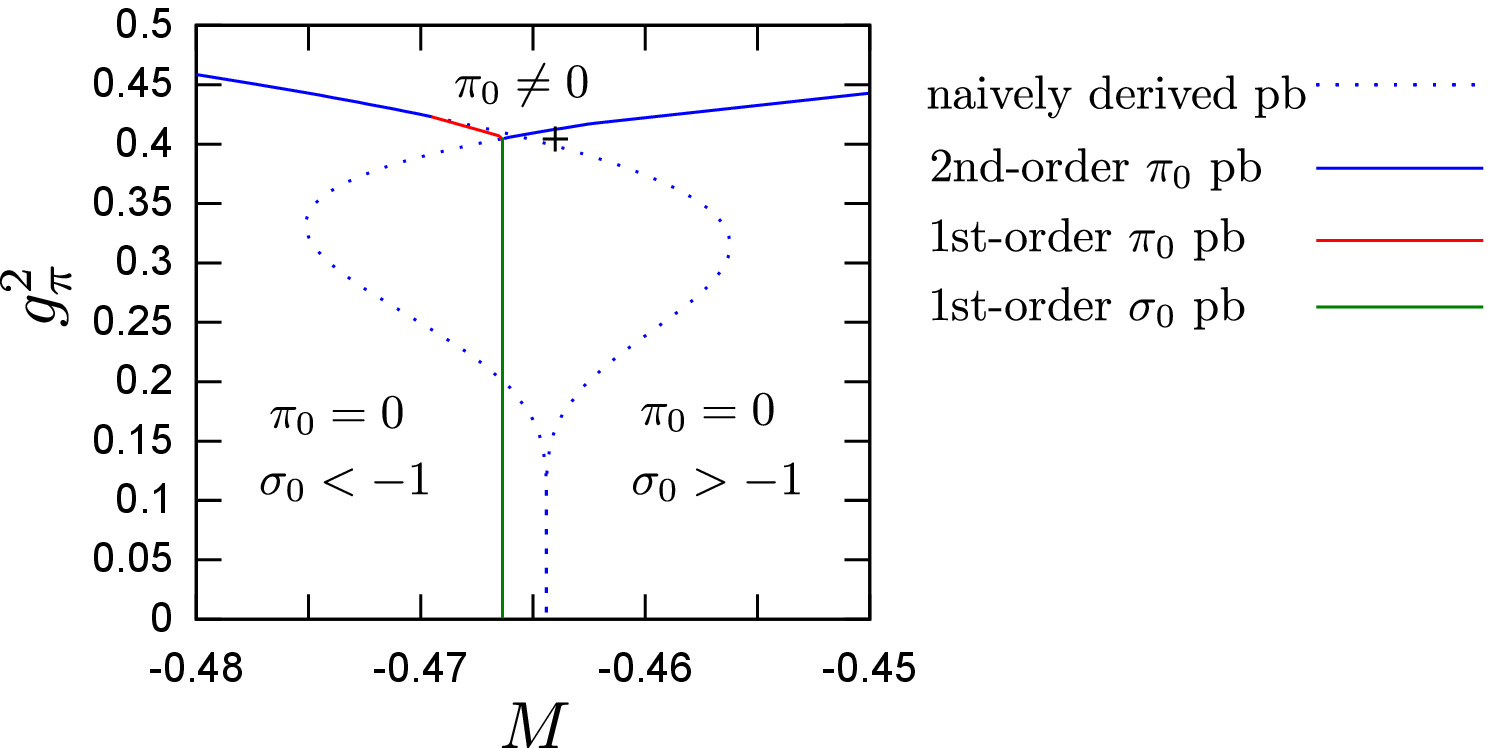} 
\caption{An expanded version of Fig.~\ref{NMgg}.
A blue dotted curve is the naively 
derived phase boundary. The true phase boundaries 
are composed of the three parts.
The fine-tuned point as a cross point is located 
slightly to the right and below the self-crossing point.}
\label{ENMgg}
\end{figure}

Let us look at these fine-tuned parameters in terms of the phase diagram. 
By this we can verify our fine-tuning yields the chiral-symmetric continuum theory. 
We first consider the non-zero value of 
$g_{\sigma}^{2}$ as $g_{\sigma}^{2}=0.6$ to reveal properties of the phase 
diagram. By hiding the lattice parameter with $a=1$ 
the fine-tuned point $(M(0.6),g_{\pi}^{2}(0.6))$ is given by
\begin{align} 
M(g_{\sigma}^{2}=0.6)\,&=\,-0.464,
\label{Mft}
\\
g_{\pi}^{2}(g_{\sigma}^{2}=0.6)\,&=\, 0.404.
\label{gpft}
\end{align}
Now we consider the $M$-$g_{\pi}^{2}$ phase diagram with $g_{\sigma}^{2}=0.6$.
According to the case of the Wilson Gross-Neveu model \cite{TNU}, 
the phase boundary has a self-crossing point and the fine-tuned point
is located slightly inside and below the self-crossing 
point in the parity symmetric phase. Besides the phase boundary naively derived 
from the gap equations no longer describes the true one near the self-crossing point, 
and we need study the effective potential to find the true critical lines including the 
1st order ones. Here we will show these situations are common with our cases.
The gap equations for the two couplings are given by
\begin{align}
M_{c}\,&=\,\sigma_{0}\Big(1-{g_{\sigma}^{2}\over{g_{\pi}^{2}}}\Big) -2g_{\sigma}^{2}\int{d^{2}k\over{(2\pi)^{2}}}
{M_{f}^{(1)}(k)\over{(\sigma_{0}+M_{f}^{(1)}(k))^{2}+s^{2}}},
\label{Ncricon1gg}
\\
{1\over{g_{\pi}^{2}}}\,&=\,2\int{d^{2}k\over{(2\pi)^{2}}}
{1\over{(\sigma_{0}+M_{f}^{(1)}(k))^{2}+s^{2}}},
\label{Ncricon2gg}
\end{align}
Here we come back to the unshifted definition of $\sigma_{0}$. 
In Fig.~\ref{NMgg} and Fig.~\ref{ENMgg} we depict the $M_{c}(g_{\pi}^{2})$ 
phase boundary derived from the gap equations
(\ref{Ncricon1gg})(\ref{Ncricon2gg}) for $g_{\sigma}^{2}=0.6$.
The latter is an expanded one near the self-crossing point with the true 
phase boundaries.
In the both figures a crosspoint stands for the fine-tuned point
without $\mathcal{O}(a)$ corrections (\ref{Mft})(\ref{gpft}).
It is located slightly to the right and below the self-crossing point 
near the second order phase boundary. We note this region is 
the parity-unbroken phase. The qualitative properties of this phase 
diagram remain toward $g_{\sigma}^{2}\to 0$ where the whole structure 
moves down to $g_{\pi}^{2}=0$ with the 1st-order boundaries disappearing. 
Here the fine-tuned point (\ref{MC1})(\ref{gC2}) gets close to the endpoint 
of the 2nd-order phase boundary at $(M,g_{\pi}^{2})\to(-1,0)$, 
which corresponds to two species $(0,0)(\pi,\pi)$.
Thus the continuum limit along this fine-tuned point yields the theory 
with chiral symmetry and two massless fermions, which leads to massless pions
as Goldstone bosons.

\begin{figure}
\centering
\includegraphics[height=5cm]{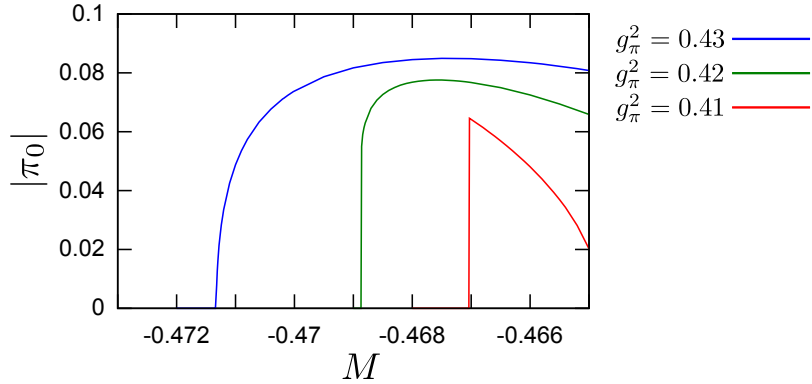} 
\caption{The order parameter $\pi_{0}$ as a function 
of $M$ for $g_{\pi}^{2}=0.41,\,0.42,\,0.43$ 
where the order of transition changes from 1st to 2nd 
in Fig.~\ref{ENMgg}.}
\label{potp1}
\end{figure}

\begin{figure}
\centering
\includegraphics[height=5cm]{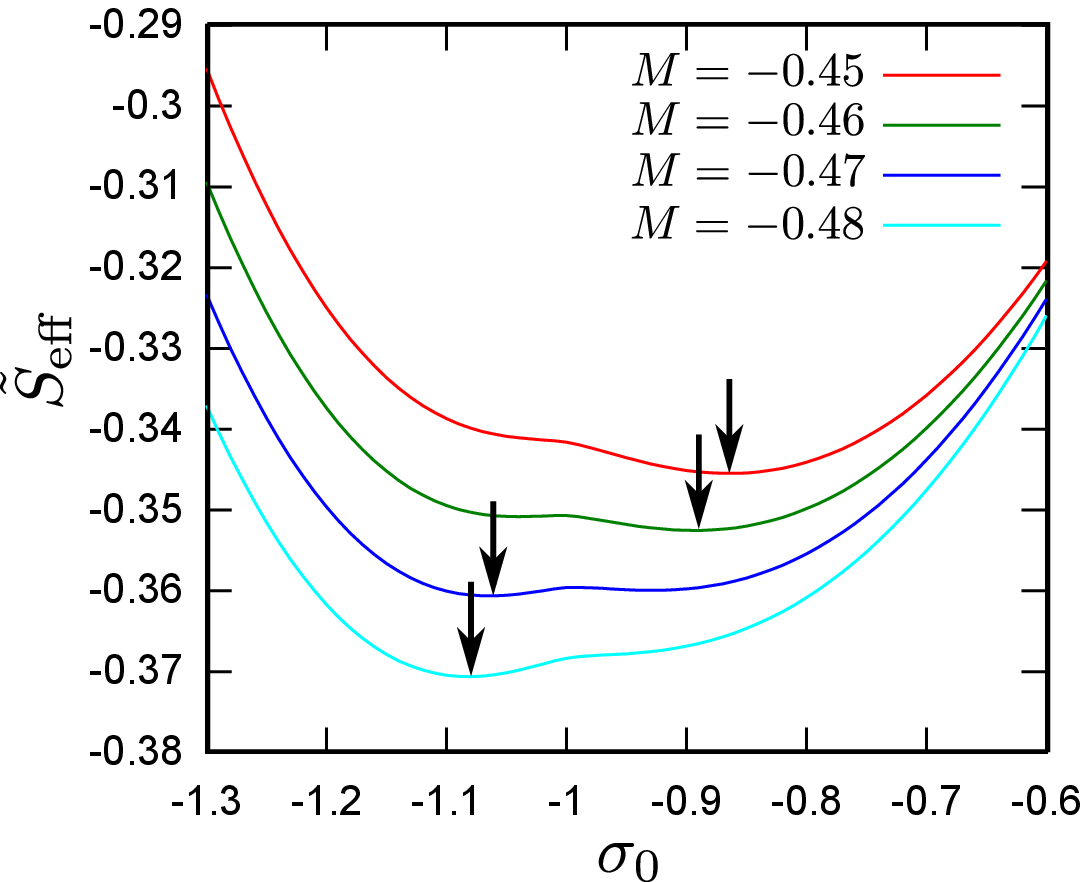} 
\caption{The $\sigma_{0}$ potential for several values of 
$M$ crossing the $\sigma_{0}$ boundary in Fig.~\ref{ENMgg}. 
The value of $\sigma_{0}$ at the minimum
changes from $\sigma_{0}>-1$ to $\sigma_{0}<-1$ 
in a form of the 1st-order transition.}
\label{pots1}
\end{figure}

Now we discuss the first order phase transition.
Although it is not essential for our purpose because 
in the limit $g_{\sigma}^{2}\to0$ the 1st-order phase boundary
disappears and the entire phase boundary becomes of 2nd order,
we can reveal other aspects of our fermions by investigating it.
As shown in \cite{TNU} there are two kinds of the 1st order phase 
boundaries in the case of Wilson fermion. One is the parity 
phase boundary, across which $\pi_{0}$ at the minimum of the effective 
potential changes from zero to nonzero. The other is related to $\sigma_{0}$, 
across which the sign of $\sigma_{0}$ at the minimum of the potential 
changes discontinuously. 
Now we will show both of them exist also in our 
case. We numerically calculate the effective potential in Eq.~(\ref{oriSeff}) 
and search the minimum of the potential. In Fig.~\ref{ENMgg} we depict the 
appearance of the 1st order phase boundaries. Here we note the true 
parity phase boundary of 2nd order as a blue solid line coincides 
with the naively derived phase boundary as a blue dotted line
at the both sides of the self-crossing. Then the 2nd-order one coming from the left 
converts to the 1st-order at some point, which is spilled out from the naively derived boundary. 
It ends at the point encountering the naively derived one again. The 1st-order phase boundary 
for $\sigma_{0}$ starts from this point, going down straight, and ends at $g_{\pi}^{2}=0$. 
In Fig.~\ref{potp1} we depict the order parameter $\pi_{0}$ as a function of $M$ 
for some fixed values of $g_{\pi}^{2}$ around which the order changes 
in Fig.~\ref{ENMgg}. Here we verify the order of the transition changes from the 2nd
to the 1st about the point. In Fig.~\ref{pots1} we depict the $\sigma_{0}$ potential for several values of 
$M$ crossing the $\sigma_{0}$ phase boundary. (Here we can take $\pi_{0}=0$ 
since it is the parity symmetric phase.) The value of $\sigma_{0}$ at the minimum
changes from $\sigma_{0}>-1$ to $\sigma_{0}<-1$ in a form of the 1st-order phase transition.
Indeed the potential describing these 1st-order transitions is also obtained 
by taking account of $\mathcal{O}(a)$ corrections.
The contribution from the correction $\delta \tilde{S}_{\rm eff}$ is given by
\begin{equation}
\delta \tilde{S}_{\rm eff}\,=\,-{8\over{3}}C_{3}\sigma_{0}^{3}+2\sigma_{0}(\sigma_{0}^2+\pi_{0}^{2})
\left( \tilde{C}_{1}+{1\over{4\pi}}\log{\sigma_{0}^{2}+\pi_{0}^{2}\over{e}} \right),
\label{correct1}
\end{equation}
with $C_{3}=-0.0923$ and $\tilde{C}_{1}=-0.0741$.
We can qualitatively reproduce the above results from the effective potential 
with these corrections. We can obtain the same but reversed phase structure 
for the right cusp by choosing $\alpha=1/a$ in (\ref{D0})(\ref{D1}). 
We also note the sign of $\sigma_{0}$ continuously changes at $M=0$. It is related with the 
discrete chiral symmetry ($\sigma_{0}\to-\sigma_{0}$) of the effective action 
(\ref{oriSeff}) for $M=0$ up to a irrelevant sign. This symmetry indicates 
interesting possibility of another continuum limit corresponding to the case 
of $\alpha=0$ in (\ref{D0})(\ref{D1}). We will discuss details on this topic 
in the end of this section.

\begin{figure}
\centering
\includegraphics[height=5cm]{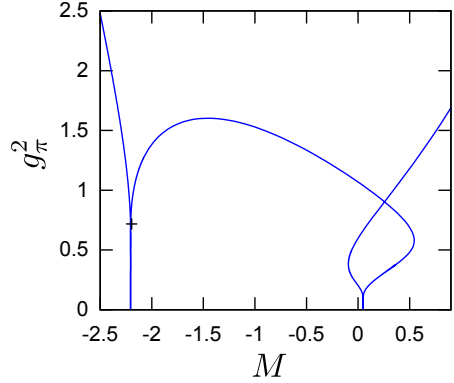} 
\caption{The naively derived phase boundary $M_{c}(g_{\pi}^{2})$ 
in the case of the naive fermion with $M_{f}^{(1)}+M_{f}^{(2)}$ 
for $g_{\sigma}^{2}=1.2$. The fine-tuned point $(-2.205, 0.720)$ 
as a crosspoint is located near the self-crossing point.}
\label{N3Mgg}
\end{figure}

\begin{figure}
\centering
\includegraphics[height=5cm]{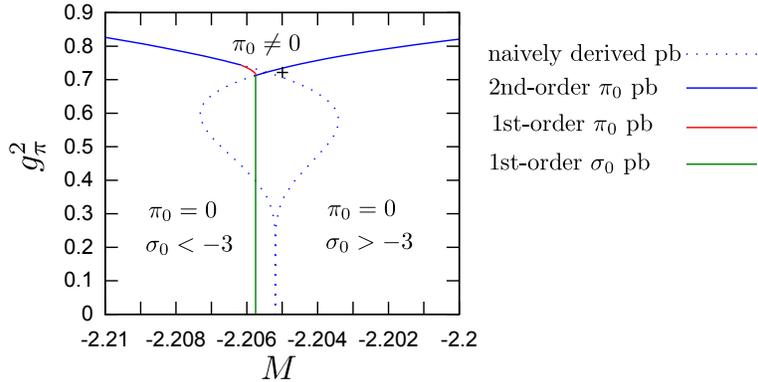} 
\caption{An expanded version of Fig.~\ref{N3Mgg}.
A blue dotted curve is the naively 
derived phase boundary. The true phase boundaries 
are composed of the three parts.
The fine-tuned point is located 
slightly to the right and below the self-crossing point.}
\label{EN3Mgg}
\end{figure}

In Fig.~\ref{N3Mgg} and \ref{EN3Mgg} we depict the corresponding figures 
for the flavored mass $M_{f}^{(1)}+M_{f}^{(2)}$. We take $g_{\sigma}^{2}=1.2$ 
to make the structure enhanced, where the fine-tuned point for the left cusp is
given by $(M,g_{\pi}^{2})=(-2.205, 0.720)$. The results are qualitatively the 
same as the previous case. In this case the continuum limit along with the 
fine-tuned point leads to the single-flavor theory with one of the species 
at $(0,0)$.

\begin{figure}
\centering
\includegraphics[height=5cm]{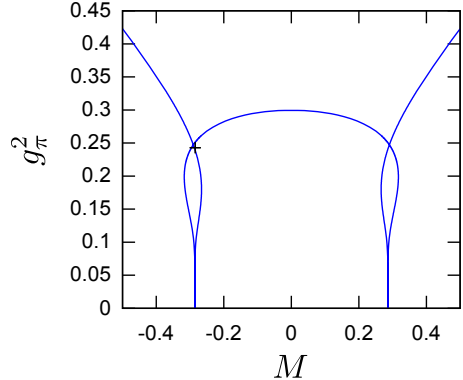} 
\caption{The naively derived phase boundary $M_{c}(g_{\pi}^{2})$ 
for the staggered fermion with the Adams-type mass 
with $g_{\sigma}^{2}=0.4$. The fine-tuned point $(-0.286, 0.243)$ 
as a crosspoint is located near the self-crossing point.}
\label{StMgg}
\end{figure}

\begin{figure}
\centering
\includegraphics[height=5cm]{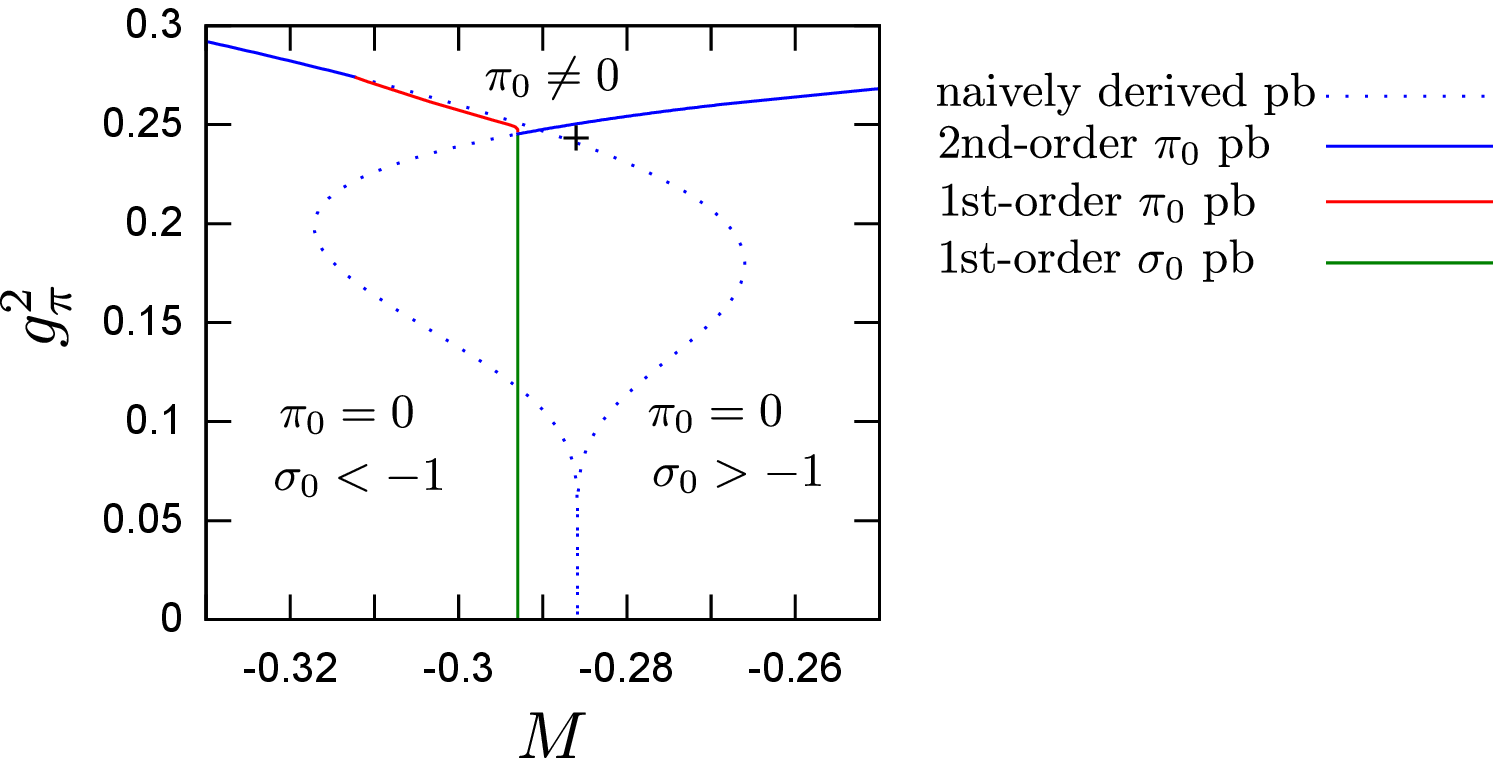} 
\caption{The expanded version of Fig.~\ref{StMgg}.
A blue dotted curve is the naively 
derived phase boundary. The true phase boundaries are 
composed of the three parts. The fine-tuned point is located 
slightly to the right and below the self-crossing point. }
\label{EStMgg}
\end{figure}

\begin{figure}
\centering
\includegraphics[height=5cm]{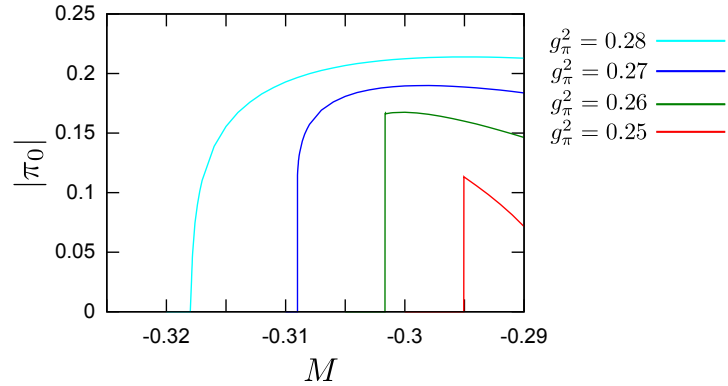} 
\caption{The order parameter $\pi_{0}$ as a function 
of $M$ for $g_{\pi}^{2}=0.25,\,0.26,\,0.27,\,0.28$ 
where the order of transition changes from 1st to 2nd 
in Fig.~\ref{EStMgg}.}
\label{potp2}
\end{figure}

\begin{figure}
\centering
\includegraphics[height=5cm]{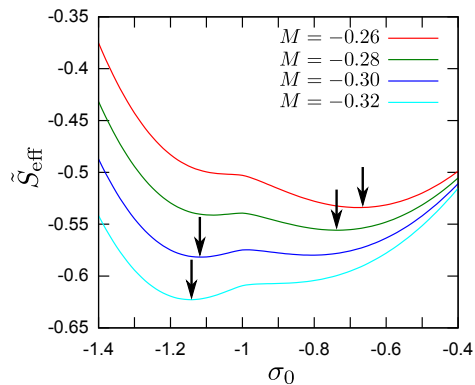} 
\caption{The $\sigma_{0}$ potential for several values of 
$M$ crossing the $\sigma_{0}$ boundary 
in Fig.~\ref{EStMgg}. The value of $\sigma_{0}$ at the minimum
changes from $\sigma_{0}>-1$ to $\sigma_{0}<-1$ 
in a form of the 1st-order transition.}
\label{pots2}
\end{figure}

We apply the same approach to the staggered Gross-Neveu model with the 
Adams-type flavored mass in Eq.~(\ref{Msta}). As seen in Eq.~(\ref{stdet}), 
the determinant in the logarithm in the effective action is given by 
the product of two determinants of the naive fermions with the mass 
$\pm M_{f}=\pm \cos (k_{1}/2)\cos (k_{2}/2)$. Thus we only have to
add the contributions from the two sectors. Here we take the 
constant $\alpha$ as $\alpha=-1/a$ and redefine 
$\sigma_{0}+1/a \,\,\,\to \,\,\,\sigma_0$ for a while. 
With this choice we can discuss the left cusp related to the taste with the positive 
flavor-chirality. Then the effective potential with the $\sigma_{0}$ shift is given by
\begin{align}
\tilde{S}_{\rm eff}(\sigma_{0},\pi_{0})
\,&=\, {(\sigma_{0}-(M+1/a))^2\over{2g_{\sigma}^{2}}}+{\pi_{0}^{2}\over{2}g_{\pi}^2}-I^{+}-I^{-},
\\
I^{\pm}&=\int^{\pi/a}_{-\pi/a}{d^{2}k\over{(2\pi)^{2}}}\log[D^{\pm}_{0}+D^{\pm}_{1}],
\\
D_{0}^{\pm}&= \sum_{\mu}{\sin ^2 {k_{\mu}a\over{2}}\over{a^2}}+\sigma_{0}^{2}+\pi_{0}^2+\Big({-1\pm\cos {k_{1}a\over{2}}\cos {k_{2}a\over{2}}\over{a}} \Big)^2 .
\\
D_{1}^{\pm}&= 2\sigma_{0}\Big({-1\pm\cos {k_{1}a\over{2}}\cos {k_{2}a\over{2}}\over{a}} \Big).
\end{align}
We expand $I$ with respect to $D_{1}/D_{0}$ as
\begin{align}
I^{\pm}&=I_{0}^{\pm}+\sum_{n=1}I_{n}^{\pm},
\\
I_{0}^{\pm}&=\int^{\pi/a}_{-\pi/a}{d^{2}k\over{(2\pi)^{2}}}\log D_{0}^{\pm},
\\
I_{n}^{\pm}&=-{(-1)^n \over{n}} \int^{\pi/a}_{-\pi/a}{d^{2}k\over{(2\pi)^{2}}} 
{(D_{1}^{\pm})^{n}\over{(D_{0}^{\pm})^{n}}}\,\,\,\,\,\,\,(n\geq 1 ),
\end{align} 
For the continuum limit $a\to 0$, only the $I_{0}^{\pm}$, $I_{1}^{\pm}$ and $I_{2}^{\pm}$ remains nonzero as in the previous case.
\begin{align} 
I_{0}^{+}+I_{0}^{-}& = \tilde{C}_{0}(\sigma_{0}^{2}+\pi_{0}^2)-{1\over{\pi}}(\sigma_{0}^{2}+\pi_{0}^2)\log{4a^{2}(\sigma_{0}^{2}+\pi_{0}^2)\over{e}}
\,\,\,\,\,\,(\tilde{C}_{0}=1.177), 
 \label{st_int1} \\
I_{1}^{+}+I_{1}^{-}& = {2\sigma_{0}\over{a}}C_{1}\,\,\,\,\,(C_{1}=-0.896),
 \label{st_int2} \\
I_{2}^{+}+I_{2}^{-}& = -2\sigma_{0}^{2}C_{2}\,\,\,\,\,(C_{2}=0.404).
  \label{st_int3}
\end{align}
Details of calculations are shown in Appendix \ref{sec:SeffC}.
The effective potential and the fine-tuned point without $\mathcal{O}(a)$ corrections
$(M(g_{\sigma}^{2}),g_{\pi}^{2}(g_{\sigma}^{2}))$ are given 
by the equations similar to Eqs.~(\ref{renS})-(\ref{renaS}) as following.
The effective potential for $a\to 0$ in this case is given by
\begin{align}
\tilde{S}_{\rm eff}&=-\Big({M+1/a\over{g_{\sigma}^2}}+{2\over{a}}C_{1}\Big)\sigma_{0}
+\Big( {1\over{2g_{\pi}^2}}-\tilde{C}_{0} + {1\over{\pi}}\log 4a^{2} \Big)\pi_{0}^{2}
\nonumber\\
&+\Big( {1\over{2g_{\sigma}^2}}-\tilde{C}_{0}+ 2C_{2} + {1\over{\pi}}\log 4a^{2} \Big)\sigma_{0}^{2}
+{1\over{\pi}}(\sigma_{0}^{2}+\pi_{0}^2)\log{\sigma_{0}^{2}+\pi_{0}^2\over{e}}.
\label{renSst}
\end{align}
Then the tuned point for the chiral limit 
without $\mathcal{O}(a)$ corrections is
\begin{align}
M&=-{2g_{\sigma}^{2}\over{a}}C_{1}-1,
\label{MC1st}
\\
g_{\pi}^{2}&={g_{\sigma}^{2}\over{4C_{2}g_{\sigma}^{2}+1}},
\label{gC2st}
\end{align}
We again introduce the scale parameter ($\Lambda$-parameter) as
\begin{align}
2a\Lambda = {\rm exp}\left[{\pi\over{2}} \tilde{C}_{0}-\pi C_{2}-{\pi\over{4g_{\sigma}^{2}}}\right].
\label{lambdast}
\end{align}
where we note the lattice spacing $a$ always appears with a factor $2$,
which is specific to the staggered fermions. 
The coupling renormalization for the chiral and continuum limit is given by
\begin{align}
{1\over{2g_{\sigma}^{2}}}&=\tilde{C}_{0}- 2C_{2} + {1\over{\pi}}\log \left( {1\over{4\Lambda^{2}a^{2}}}\right),
\label{arengsst}
\\
{1\over{2g_{\pi}^{2}}}&=\tilde{C}_{0} + {1\over{\pi}}\log \left( {1\over{4\Lambda^{2}a^{2}}}\right).
\label{arengpst}
\end{align}
where we keep $\Lambda$ finite when taking the continuum limit $a\to 0$.
Finally the renormalized effective potential in the chiral and continuum limit is given by
\begin{equation}
\tilde{S}_{\rm eff}={1\over{\pi}}(\sigma_{0}^{2}+\pi_{0}^{2})\log{\sigma_{0}^{2}+\pi_{0}^{2}\over{e\Lambda^{2}}}
\label{renaSst}
\end{equation}
In this case we take $g_{\sigma}^{2}=0.4$ as an example, 
then the fine-tuned point is given by 
\begin{align} 
M(g_{\sigma}^{2}=0.4)\,&=\,-0.286,
\label{Mftst}
\\
g_{\pi}^{2}(g_{\sigma}^{2}=0.4)\,&=\, 0.243.
\label{gpftst}
\end{align}
The gap equations in this case are given by
\begin{align}
M_{c}\,&=\,\sigma_{0}\Big(1-{g_{\sigma}^{2}\over{g_{\pi}^{2}}}\Big) +8g_{\sigma}^{2}\sigma_{0}
\int{dk^{2}\over{(2\pi)^2}}{c_{1}^{2}c_{2}^{2}
\over{((\sigma_{0}+c_{1}c_{2})^{2}+\pi_{0}^2+s^{2})((\sigma_{0}-c_{1}c_{2})^2+\pi_{0}^2+s^{2})}},
\label{Stcricon1gg}
\\
{1\over{g_{\pi}^{2}}}\,&=\,4\int{dk^{2}\over{(2\pi)^2}}{\sigma_{0}^{2}+s^{2}+c_{1}^{2}c_{2}^{2}
\over{((\sigma_{0}+c_{1}c_{2})^{2}+\pi_{0}^2+s^{2})((\sigma_{0}-c_{1}c_{2})^2+\pi_{0}^2+s^{2})}}.
\label{Stcricon2gg}
\end{align}
Here we come back to the unshifted definition of $\sigma_{0}$. 
In Fig.~\ref{StMgg} and Fig.~\ref{EStMgg} we depict the phase boundary $M(g_{\pi}^{2})$
naively derived from the above gap equations for $g_{\sigma}^{2}=0.4$.
The latter is an expanded one near the self-crossing point with the true phase boundaries also depicted.
The fine-tuned point (\ref{Mftst})(\ref{gpftst}) is located slightly to the right and below 
the self-crossing point near the true second order phase boundary in the parity symmetric phase.
Toward the week-coupling limit $g_{\sigma}^{2}\to 0$ the phase structure moves down to $g_{\pi}^{2}=0$,
where the fine-tuned point gets close to $(M,g_{\pi}^{2})\to(-1,0)$ from the parity symmetric phase.
It means our fine-tuned point leads to the continuum theory with the chiral symmetry 
and one massless fermion corresponding to the taste with positive flavor-chirality.
The situation about the first order phase boundary is the same as the naive case. 
In Fig.~\ref{EStMgg} we depict the true phase boundaries for this case. 
In Fig.~\ref{potp2} we depict the order parameter $\pi_{0}$ as a function of $M$. 
Here the order of the transition changes from the 2nd to the 1st around the order-changing point. 
In Fig.~\ref{pots2} we depict the $\sigma_{0}$ potential for several values of 
$M$ crossing the $\sigma_{0}$ phase boundary. The value of $\sigma_{0}$ at the minimum
changes from $\sigma_{0}>-1$ to $\sigma_{0}<-1$ in a form of the 1st-order phase transition.

We have shown that the chirally-symmetric continuum limit can be taken by fine-tuning 
a mass parameter and two coupling constants both for the naive and staggered cases. 
It indicates we obtain the two-flavor or one-flavor massless fermions in the chiral limit 
by tuning only a mass parameter when we introduce the Adams-type \cite{Adams2} 
or Hoelbling-type \cite{Hoel} flavored masses to the $d=4$ QCD with staggered fermions. 
The less numerical expense in the staggered fermion could make 
the QCD simulations with these fermions faster than Wilson fermion. 
We need further investigation to answer this question.

Now we comment on the case that we take $\alpha=0$ in (\ref{D0})(\ref{D1}), 
which corresponds to neither of the cusps but reflects the symmetries of 
the effective potential. 
At this point the coupling is not going to
zero, and thus it is unclear how it is related to the continuum
Gross-Neveu model. However it does seem to be possible to restore
chiral symmetry there and have a divergent correlation length.  As
such it seems related to a quite special continuum limit.
Since the $M=0$ effective potentials
for the naive with $M_{f}^{(1)}$ and the staggered fermions 
with the Adams-type mass possess the $Z_{2}$ discrete chiral 
symmetry ($\sigma_{0}\to-\sigma_{0}$) up to a irrelevant sign, the renormalization 
in the linear $\sigma_{0}$ term is prohibited. Actually we have checked $C_{1}$ 
in the effective potentials as (\ref{renS}) is zero for both cases with $\alpha=0$. 
This is because 
the continuum chiral symmetry is broken while the discrete one is unbroken 
by these flavored masses. Thus, if we start with $M=0$, it appears we need 
not fine-tune the mass parameter for the massless continuum limit with the 
chiral symmetry. It indicates a strange possibility that the chirally 
symmetric continuum limit of the $d=4$ QCD with these fermions is taken 
without fine-tuning due to this symmetry. 
This strange situation can occur for any flavored mass with the discrete chiral 
symmetry up to a trivial sign such as $M_{f}=\sum_{\mu}\cos k_{\mu}$. 
However the question is whether the continuum limit stands for 
physically relevant theories. Indeed it is unlikely since the line 
$M=0$ is located at the same distance from the two cusps thus 
the continuum limit along it would have no physical fermions,
although there might exist some relevant theory without fermions like 
the Ising theory. On the other hand, in the naive fermion with 
$M_{f}^{(2)}$ or the $d=4$ staggered fermion with 
the mass proposed by Hoelbling in \cite{Hoel}, the $M=0$ line has a cusp 
in the weak coupling region. The effective actions in these cases also have 
the discrete chiral symmetry and the same situation occurs. 
Thus, the continuum limit without fine-tuning in them may 
lead to the relevant theories with the parity symmetry being broken 
since the continuum limit is taken from the Aoki phase in these cases.
This kind of the parity or CP broken theory with massless fermions would 
belong to the same universality class as minimally doubled fermions 
\cite{MD1, MD2} or the two-flavor QCD with the sign of mass being different 
between the two flavors \cite{CreutzCP}.
Further study on this topic is devoted to the future work.

%%%%%%%%%%   Summary and Discussion   %%%%%%%%%%

\section{Summary and Discussion}
\label{sec:SD}

In this paper we investigate the parity-broken phase structure for 
naive and staggered fermions with the flavored mass by using the 
two-dimensional lattice Gross-Neveu models. We have shown the Aoki 
phase exists both in staggered and naive cases reflecting the 
mass splitting in species. 

In Sec.~\ref{sec:NGN} we study the phase structure for the naive
Gross-Neveu model with the flavored masses. We consider the two types
of flavored mass terms for 2d naive fermions, which cause two
different kinds of mass splitting in species. We also consider a
linear combination of these terms. We solve the gap equations for the
large $N$ limit and obtain the second order phase boundaries in the
$M$-$g^{2}$ plane.  The parity broken phase diagram has some common
properties with the Wilson case, and reflects the mass splitting.
We can make varieties of phase structures depending on
arbitrary linear combinations of the two types of the masses.  In
Sec.~\ref{sec:SGN} we consider the generalized staggered Gross-Neveu
model including two types of four-point interactions.  We take the
same process as in the case of the naive fermion to obtain the phase
diagram for the staggered fermion with the Adams-type flavored mass.
We show the Aoki phase exists also in this case reflecting the mass
splitting of tastes.  This elucidation can contribute to the practical
application of these fermions and their overlap versions.  In
Sec.~\ref{sec:CL} we discuss the continuum limit of these Gross-Neveu
models around the cusps in the phase digram.  We show that the
chirally-symmetric continuum limit with the number of massless species
associated with each of the cusps can be taken by fine-tuning a mass parameter
and two coupling constants in both cases. From this we speculate the chiral 
limit can be taken by fine-tuning only a mass parameter
in $d=4$ lattice QCD with staggered fermions with the Adams-type \cite{Adams2} 
or Hoelbling-type \cite{Hoel} masses. It indicates we can obtain the one- 
or two-flavor massless fermions in the continuum from the staggered setup 
and regard massless pions as Goldstone bosons due to the spontaneous
chiral symmetry breaking as in the case with Wilson fermion.  
These approaches avoid the use of the rooting approximation to reduce 
the number of tastes.  We also study the first order phase boundaries 
peculiar to the two-coupling cases of the lattice Gross-Neveu models. 
We show there exist two kinds of the first order phase boundaries with 
respect to parity and chiral symmetry breaking as in the case of Wilson fermion.

We comment on the possible advances of the one-flavor or two-flavor
staggered fermions without rooting discussed in this paper compared to
Wilson fermion.  Taking account of less numerical expense in the
staggered fermion, it will be numerically better than Wilson fermion
in the lattice QCD simulations.  We can estimate how good it is easily
by calculating simple examples.  Future works will be devoted to this
study.

%%%%%%%%%%   ACKNOWLEDGMENTS   %%%%%%%%%%

\begin{acknowledgments}
MC is grateful to the Alexander von Humboldt Foundation for support 
for visits to the University of Mainz.  This manuscript has been 
authored under contract number DE-AC02-98CH10886 with the 
U.S.~Department of Energy. Accordingly, the U.S. Government retains
a non-exclusive, royalty-free license to publish or reproduce 
the published form of this contribution, or allow others to do so, 
for U.S.~Government purposes. TK is supported by the JSPS Institutional 
Program for Young Researcher Overseas Visits. TM is grateful to Taku Izubuchi
for fruitful discussion and hearty encouragement. TM is supported by 
Grand-in-Aid for the Japan Society for Promotion of Science (JSPS) 
Research Fellows(No.\ 21-1226).

\end{acknowledgments}

%%%%%%%%%%   APPENDIX   %%%%%%%%%%

\appendix

\section{Derivation of the effective potentials}

In this appendix we evaluate the integrals which are required for the
effective potentials for the cases with the naive and staggered fermions.

\subsection{Naive fermion}
\label{sec:NeffC}

We have to evaluate the integrals of (\ref{naive_int0}) and
(\ref{naive_int1}) to obtain the effective potential of the model with
the naive fermion.
Let us first study the following integral,
\begin{eqnarray}
 I_0 & = & \int_{-\pi/a}^{\pi/a} \frac{d^2 k}{(2\pi)^2}
  \log\left[
       \frac{s^2}{a^2} + \sigma^2_0 + \pi^2_0
       + \left(\frac{-1+M_f}{a}\right)^2
      \right],
\end{eqnarray}
where we denote $s^2 = \sum_\mu \sin^2 (k_\mu a)$ and $M_f = \cos (k_1 a)
\cos (k_2 a)$.
If we omit a constant term which is not involving $\sigma_0$ and
$\pi_0$, it can be rewritten in an integral representation as
\begin{eqnarray}
 I_0 & \simeq & \int_0^{\sigma^2_0+\pi^2_0} d\rho\ F_0(\rho),
  \label{naive_appndx1}
  \\
 F_0(\rho) & = & \int_{-\pi/a}^{\pi/a} \frac{d^2 k}{(2\pi)^2}
  \frac{1}{s^2/a^2 + (-1+M_f)^2/a^2 + \rho}.
\end{eqnarray}
We pick up the divergent part in the limit of $a \to 0$,
\begin{eqnarray}
 F_0(\rho) & \stackrel{a \to 0}{\longrightarrow} & 
  \int_{-\pi/(2a)}^{3\pi/(2a)} \frac{d^2 k}{(2\pi)^2}
  \left(
   \frac{1}{\sum_\mu k_\mu^2 + \rho}
   +    \frac{1}{\sum_\mu (k_\mu-\pi)^2 + \rho}
  \right) + c_0,
  \\
 c_0 & = & \int_{-\pi/2}^{3\pi/2} \frac{d^2 \xi}{(2\pi)^2}
  \left(
   \frac{1}{s^2 + (-1+M_f)^2}
  - \frac{1}{\sum_\mu \xi_\mu^2} 
  - \frac{1}{\sum_\mu (\xi_\mu-\pi)^2} 
  \right)
  \left(= 0.0421\right).
\end{eqnarray}
Here we shift the Brillouin zone to treat the divergent part, which
originates from two massless modes around $k = (0,0)$ and $(\pi, \pi)$.
We then find the following expression by comparing the first term with the
corresponding integral in the continuum theory,
\begin{equation}
  \int_{-\pi/(2a)}^{3\pi/(2a)} \frac{d^2 k}{(2\pi)^2}
  \left(
   \frac{1}{\sum_\mu k_\mu^2 + \rho}
   +    \frac{1}{\sum_\mu (k_\mu-\pi)^2 + \rho}
  \right)
  = \frac{1}{2\pi} \log \frac{1}{a^2 \rho} + c_0' 
  \quad \left(c_0' = 0.325\right).
\end{equation}
Therefore the integral is given by
\begin{equation}
 F_0(\rho) = \frac{1}{2\pi} \log \frac{1}{a^2 \rho} + \tilde C_0 
  \quad \left(\tilde C_0 = 0.367\right),
\end{equation}
where $\tilde C_0 = c_0 + c_0'$ is the constant used in
(\ref{naive_int2}).
By substituting this into (\ref{naive_appndx1}), we obtain the
expression in (\ref{naive_int2})
\begin{equation}
 I_0 (a \to 0) = \tilde C (\sigma_0^2 + \pi_0^2) - \frac{1}{2\pi} 
  (\sigma_0^2 + \pi_0^2) \log \frac{a^2(\sigma_0^2+\pi_0^2)}{e}.
\end{equation}

Next we show the integral expressions of (\ref{naive_int3}) and
(\ref{naive_int4}).
They are given by 
\begin{eqnarray}
 C_1 & = & \int_{-\pi}^{\pi} \frac{d^2 \xi}{(2\pi)^2}
  \frac{-1 + M_f}{s^2 + (-1 + M_f)^2} \quad ( = -0.446 ), \\
 C_2 & = & \int_{-\pi}^{\pi} \frac{d^2 \xi}{(2\pi)^2}
  \left(\frac{-1 + M_f}{s^2 + (-1 + M_f)^2}\right)^2 \quad ( = 0.201 ).
\end{eqnarray}
These integrals are sufficient to consider the continuum limit of the
model, but not to discuss the 1st-order phase transition.
The $\mathcal{O}(a)$ corrections come from the following integrals,
\begin{equation}
 I_3(a \to 0) = \frac{8}{3} \sigma_0^3 a C_3, \quad
 C_3 = \int_{-\pi}^{\pi} \frac{d^2 \xi}{(2\pi)^2}
  \left(\frac{-1 + M_f}{s^2 + (-1 + M_f)^2}\right)^3 \quad ( = - 0.0923 ),
\end{equation}
\begin{eqnarray}
 \delta I_1 & = & I_1 - \frac{2\sigma_0}{a} C_1
  \nonumber \\
 & = & 2 \sigma_0 \int_{-\pi/a}^{\pi/a} \frac{d^2 k}{(2\pi)^2}
  \left(
   \frac{(-1+M_f)/a}{s^2/a^2+(-1+M_f)^2/a^2+\sigma_0^2+\pi^2_0}
   -    \frac{(-1+M_f)/a}{s^2/a^2+(-1+M_f)^2/a^2}
  \right)
  \nonumber \\
 & = & - 2 \sigma_0 a \int_0^{\sigma_0^2+\pi_0^2} d \rho \ F_1(\rho) ,
   \label{naive_appndx2} \\
 F_1(\rho) & = & \frac{1}{a} \int_{-\pi/a}^{\pi/a} \frac{d^2 k}{(2\pi)^2}
  \frac{(-1+M_f)/a}{(s^2/a^2+(-1+M_f)^2/a^2+\rho)^2}.
\end{eqnarray}
We can evaluate the second one in a similar way by splitting into a
divergent part and a finite constant,
\begin{eqnarray}
 F_1(\rho) & \stackrel{a \to 0}{\longrightarrow} &
   - \frac{1}{2} \int_{-\pi/(2a)}^{3\pi/(2a)} \frac{d^2 k}{(2\pi)^2}
  \left(
   \frac{\sum_\mu k_\mu^2}{\left(\sum_\mu k_\mu^2 + \rho\right)^2}
   + \frac{\sum_\mu (k_\mu-\pi)^2}{\left(\sum_\mu (k_\mu-\pi)^2 + \rho\right)^2}
  \right) + c_1 \\
 c_1 & = & \int_{-\pi/2}^{3\pi/2} \frac{d^2 \xi}{(2\pi)^2}
  \left(
   \frac{(-1+M_f)}{(s^2 + (-1+M_f)^2)^2}
  + \frac{1}{2\sum_\mu \xi_\mu^2} 
  + \frac{1}{2\sum_\mu (\xi_\mu-\pi)^2} 
  \right)
  \left(= 0.00912\right).
  \nonumber \\
\end{eqnarray}
The divergent part is given by
\begin{equation}
 \int_{-\pi/(2a)}^{3\pi/(2a)} \frac{d^2 k}{(2\pi)^2}
  \left(
   \frac{\sum_\mu k_\mu^2}{\left(\sum_\mu k_\mu^2 + \rho\right)^2}
   + \frac{\sum_\mu (k_\mu-\pi)^2}{\left(\sum_\mu (k_\mu-\pi)^2 +
				    \rho\right)^2}
  \right)
  = \frac{1}{2\pi} \log \frac{1}{a^2 \rho} + c_1'
  \quad ( c_1' = 0.166).
\end{equation}
Thus we obtain
\begin{equation}
 F_1(\rho) = - \frac{1}{4\pi} \log \frac{1}{a^2 \rho} + \tilde C_1
  \quad \left(\tilde C_1 = c_1 - \frac{c_1'}{2} = -0.0741\right).
\end{equation}
By substituting this expression into (\ref{naive_appndx2}), we obtain
\begin{equation}
 \delta I_1 = - 2 \sigma_0 a
  \left[
   \tilde C_1 (\sigma_0^2 + \pi_0^2) + \frac{1}{4\pi} (\sigma_0^2 + \pi_0^2)
   \log \frac{a^2(\sigma_0^2 + \pi_0^2)}{e}
  \right].
\end{equation}
These integrals contribute to the $\mathcal{O}{(a)}$ corrections to the
effective potential (\ref{correct1}).

\subsection{Staggered fermion}
\label{sec:SeffC}

We evaluate the integrals required for the effective potentials with the
staggered fermion.
Explicit forms of the finite constants in (\ref{st_int2}) and
(\ref{st_int3}) are simply given by
\begin{eqnarray}
 C_1 & = & \int_{-\pi}^\pi \frac{d^2 \xi}{(2\pi)^2}
  \left[
   \frac{-1+M_f}{s^2 + (-1+M_f)} + \frac{-1-M_f}{s^2 + (-1-M_f)}
  \right]
  \quad (= - 0.896),
  \\
 C_2 & = & \int_{-\pi}^\pi \frac{d^2 \xi}{(2\pi)^2}
  \left[
   \left(\frac{-1+M_f}{s^2 + (-1+M_f)}\right)^2
   + \left(\frac{-1-M_f}{s^2 + (-1-M_f)}\right)^2
  \right]
  \quad (= 0.404),
\end{eqnarray}
where we use similar symbols as the naive fermion case, $s^2 =
\sum_\mu \sin^2 (k_\mu a/2)$, $M_f = \cos (k_1 a/2) \cos (k_2 a/2)$.

The integral (\ref{st_int1}) is slightly complicated,
but can be evaluated in a similar manner.
Omitting a constant term independent on $\sigma_0$ and $\pi_0$, it can
be written as
\begin{eqnarray}
 I_0^+ & = & \int_{-\pi/a}^{\pi/a} \frac{d^2 k}{(2\pi)^2}
  \log \left[
	\frac{s^2}{a^2} + \sigma_0^2 + \pi_0^2
	+ \left(\frac{-1+M_f}{a}\right)^2
       \right]
  \nonumber \\
  & \simeq & \int_0^{\sigma_0^2 + \pi_0^2} d \rho\ F(\rho), \\
 F(\rho) & = & \int_{-\pi/a}^{\pi/a} \frac{d^2 k}{(2\pi)^2}
  \frac{1}{s^2/a^2 + (-1+M_f)^2/a^2 + \rho}.
\end{eqnarray}
We can split this integral into a divergent part and a finite constant
in the limit of $a \to 0$,
\begin{eqnarray}
 F(\rho) & \stackrel{a \to 0}{\longrightarrow} &
  \int_{-\pi/a}^{\pi/a} \frac{d^2 k}{(2\pi)^2} 
  \frac{1}{\sum_\mu k_\mu^2/4 + \rho} + c_0^+,
  \nonumber \\
 c_0^+ & = & \int_{-\pi}^{\pi} \frac{d^2 \xi}{(2\pi)^2} 
  \left(
   \frac{1}{s^2+(-1+M_f)^2} - \frac{1}{\sum_\mu \xi_\mu^2/4}
  \right)
  \quad \left( = 0.0440\right).
\end{eqnarray}
The divergent part is given by
\begin{equation}
 \int_{-\pi/a}^{\pi/a} \frac{d^2 k}{(2\pi)^2} 
  \frac{1}{\sum_\mu k_\mu^2/4 + \rho}
   = \frac{1}{\pi} \log \frac{1}{4a^2\rho} + {C_0^+}'
   \quad \left( {C_0^+}' = 0.798\right).
\end{equation}
Thus we obtain
\begin{equation}
 F(\rho) = \frac{1}{\pi} \log \frac{1}{4a^2\rho} + \tilde C_0^+
  \quad \left( \tilde C_0^+ =  C_0^+ + {C_0^+}' = 0.842 \right).
\end{equation}
The corresponding integral becomes
\begin{equation}
 I_0^+(a \to 0) = \tilde C_0^+ (\sigma_0^2 + \pi_0^2)
  - \frac{1}{\pi}(\sigma_0^2 + \pi_0^2) 
  \log \frac{4a^2(\sigma_0^2 + \pi_0^2)}{e}.
\end{equation}

The other integral is written as
\begin{eqnarray}
 I_0^- &\simeq& C_0^- (\sigma_0^2 + \pi_0^2) + \mathcal{O}(a), \\
 C_0^- & = & \int_{-\pi}^{\pi} \frac{d^2 \xi}{(2\pi)^2}
  \frac{1}{s^2 + (1+M_f)^2}
   \quad ( = 0.333)
\end{eqnarray}
where we again omit a constant independent on $\sigma_0$ and $\pi_0$.
As a result we obtain the expression of (\ref{st_int1}),
\begin{equation}
 I_0^+ + I_0^- = \tilde C_0 (\sigma_0^2 + \pi_0^2)
  - \frac{1}{\pi}(\sigma_0^2 + \pi_0^2) 
  \log \frac{4a^2(\sigma_0^2 + \pi_0^2)}{e}
  \quad \left( \tilde C_0 = \tilde C_0^+ + C_0^- = 1.177\right).
\end{equation}

%%%%%%%%%%   References   %%%%%%%%%%

\end{document}